\documentclass[journal]{IEEEtran}

\usepackage{amsmath, amssymb, bm, cite, epsfig, psfrag, mathtools,amsfonts}
\usepackage{epstopdf}
\usepackage{graphicx}
\usepackage{algorithm,algpseudocode}
\usepackage{array}
\usepackage[margin=10pt,font=small,labelfont=bf]{caption}
\usepackage{bbm}
\usepackage{multirow}
\usepackage[usenames,dvipsnames]{xcolor}

\usepackage{subfigure}
\usepackage{etoolbox}
\usepackage{pbox}
\usepackage{threeparttable}

\usepackage[normalem]{ulem}

\newtoggle{onecol}
\togglefalse{onecol} 
\graphicspath{{figures/}}
\DeclareGraphicsExtensions{.pdf,.png,.jpg}

\def\beq{\begin{equation}}
\def\eeq{\end{equation}}
\def\beqa{\begin{eqnarray}}
\def\eeqa{\end{eqnarray}}
\def\beqan{\begin{eqnarray*}}
\def\eeqan{\end{eqnarray*}}

\def\C{{\mathbb{C}}}

\def\x{\times}

\def\gbar{\overline{g}}

\def\SNR{\mbox{\small \sffamily SNR}}

\def\Exp{\mathbb{E}}

\def\tm1{t\! - \! 1}
\def\tp1{t\! + \! 1}

\def\ubf{\mathbf{u}}

\def\vbf{\mathbf{v}}

\def\Hbf{\mathbf{H}}

\def\Pbf{\mathbf{P}}

\def\Qbf{\mathbf{Q}}

\def\Vbf{\mathbf{V}}

\def\LOS{\text{LOS}}
\def\NLOS{\text{NLOS}}
\def\outage{\text{out}}
\def\out{\text{out}}
\def\los{\text{los}}

\begin{document}
\bibliographystyle{IEEEtran}

\title{Millimeter Wave Channel Modeling and Cellular Capacity Evaluation}

\author{
    Mustafa Riza Akdeniz,~\IEEEmembership{Student~Member,~IEEE},
    Yuanpeng Liu,~\IEEEmembership{Student~Member,~IEEE},
    Mathew K. Samimi,~\IEEEmembership{Student~Member,~IEEE},
    Shu Sun,~\IEEEmembership{Student~Member,~IEEE},
    Sundeep Rangan,~\IEEEmembership{Senior Member,~IEEE},
    Theodore S. Rappaport,~\IEEEmembership{Fellow,~IEEE},
    Elza Erkip,~\IEEEmembership{Fellow,~IEEE}
    \thanks{This material is based upon work supported by the National Science
    Foundation under Grants No. 1116589 and 1237821 as well as generous support
    from Samsung, Nokia Siemens Networks and InterDigital Communications.}
    \thanks{M. Akdeniz (email:makden01@students.poly.edu),
           Y. Liu (email:yliu20@students.poly.edu),
           M. Samimi (email:mathewsamimi@gmail.com),
           S. Sun (email:ss7152@nyu.edu),
           S. Rangan (email: srangan@poly.edu),
           T. S. Rappaport (email: tsr@nyu.edu) and
           E. Erkip (email: elza@poly.edu) are with NYU WIRELESS Center,
           Polytechnic Institute of New York University, Brooklyn, NY.}
}

\maketitle

\begin{abstract}
With the severe spectrum shortage in conventional cellular bands,
millimeter wave (mmW) frequencies between 30 and 300~GHz have been
attracting growing attention as a possible candidate for next-generation
micro- and picocellular wireless networks.
The mmW bands offer orders of magnitude greater spectrum
than current cellular allocations and enable very high-dimensional
antenna arrays for further gains via beamforming and spatial multiplexing.
This paper uses recent real-world measurements at 28 and 73~GHz in New York City
to derive detailed spatial statistical models of the channels and uses
these models to provide a realistic assessment of mmW micro- and
picocellular networks in a dense urban deployment.
Statistical models are derived for key channel parameters including
the path loss, number of spatial clusters, angular dispersion and outage.
It is found that, even in highly non-line-of-sight environments,
strong signals can be detected 100m to 200m
from potential cell sites, potentially with multiple clusters to support
spatial multiplexing.
Moreover, a system simulation based on the models predicts that
mmW systems
can offer an order of magnitude increase in capacity
over current state-of-the-art 4G cellular networks with no increase in cell density
from current urban deployments.
\end{abstract}


\section{Introduction}
\label{sec:intro}

The remarkable success of cellular wireless technologies have
led to an insatiable demand for mobile data \cite{CiscoVNI:latest,EricssonMDT:latest}.
The UMTS traffic forecasts \cite{UMTSForecast}, for example, predicts that
by 2020, \emph{daily} mobile traffic
will exceed 800 MB per subscriber leading to
130 exabits ($10^{18}$) of data per year for some operators.
Keeping pace with this
demand will require new technologies
that can offer orders of magnitude
increases in cellular capacity.

To address this challenge, there has been growing interest in
cellular systems based in the so-called \emph{millimeter-wave} (mmW) bands,
between 30 and 300 GHz, where the available bandwidths are much wider than today's
cellular networks~\cite{KhanPi:11,KhanPi:11-CommMag,Ted:60Gstate11,PietBRPC:12,BocHLMP:14,RanRapE:14}.
The available spectrum at these frequencies can be easily 200 times
greater than all cellular allocations today that are currently largely constrained
to the
prime RF real estate under 3~GHz~\cite{KhanPi:11-CommMag}.
Moreover, the very small wavelengths of mmW signals combined with advances
in low-power CMOS RF
circuits enable large numbers ($\geq$ 32 elements) of  miniaturized antennas
to be placed in small dimensions.  These multiple antenna systems
can be used to form very high gain, electrically steerable arrays,
fabricated at the base station, in the skin of a cellphone,
or even within a chip \cite{Doan:04,Doan:05,ZhaLiu:09,gutierrez2009chip,Nsenga:10,Ted:60Gstate11,Rajagopal:mmWMobile,Huang:2008:MWA:1524107,Rusek:13}.
Given the very wide bandwidths and large numbers of spatial degrees of freedom,
it has been speculated that
mmW bands will play a significant role in Beyond 4G and 5G cellular systems~\cite{BocHLMP:14}.

However, the development of cellular networks in the mmW bands
faces significant technical obstacles and the precise value of mmW systems
needs careful assessment.
The increase in omnidirectional free space path loss with higher frequencies due to Friis' Law
\cite{Rappaport:02}, can be more than compensated by
a proportional increase in antenna gain with appropriate beamforming.  We will, in fact,
confirm this property experimentally below.
However, a more significant concern is that mmW signals can be severely vulnerable to
shadowing resulting in outages, rapidly varying channel conditions and intermittent connectivity.
This issue is particularly concerning in cluttered, urban deployments
where coverage frequently requires non-line-of-sight (NLOS) links.

In this paper, we use the measurements of mmW outdoor cellular
propagation
\cite{Rappaport:12-28G,Rappaport:28NYCPenetrationLoss,Samimi:AoAD,rappaportmillimeter,Rappaport:13-smallCell,Sun-Beam:13}
in 28 and 73~GHz in New York City
to derive detailed the first statistical channel models
that can be used for proper mmW system evaluation.  The models are used
to provide an initial assessment of the potential system capacity
and outage.
The NYC location was selected since it
is representative of likely initial deployments of mmW cellular
systems due to the high user density.  In addition, the urban canyon environment
provides a challenging test case for these systems due to the difficulty
in establishing line-of-sight (LOS) links -- a key concern for mmW cellular.

Although our earlier work has presented some initial analysis of the data
in \cite{Rappaport:12-28G,Rappaport:28NYCPenetrationLoss,Samimi:AoAD,rappaportmillimeter,Rappaport:13-smallCell},
this work provides much more detailed modeling necessary for cellular system
evaluation.
In particular, we develop detailed models for the spatial characteristics
of the channel and outage probabilities.
To obtain these models, several we present new data analysis techniques.
In particular, we propose a clustering algorithm
that identifies the group of paths in the angular domain from subsampled spatial
measurements. The clustering algorithm is based on a $K$-means method
with additional heuristics to determine the number of clusters. Statistical models are then derived for key cluster parameters including the number of clusters, cluster angular spread and path loss. For the inter-cluster power fractions, we propose a probabilistic model with maximum likelihood (ML) parameter estimation.
In addition, while standard 3GPP models such as \cite{3GPP36.814,ITU-M.2134}
use probabilistic LOS-NLOS models, we propose to add a third state
to explicitly model the  models the possibility of outages.

The key findings from these models are as follows:
\begin{itemize}
\item The omnidirectional
path loss is approximately 20 to 25~dB higher in the mmW frequencies relative to current
cellular frequencies
in distances relevant for small cells.  However, due to the reduced wavelength,
this loss can be completely compensated by a proportional increase in antenna gain
with no increase in physical antenna size.  Thus, with appropriate beamforming,
\emph{locations that are not in outage will not experience any effective increase in path loss and, in fact, the path loss may be decreased}.

\item Our measurements indicate that at many locations, energy arrives in clusters from multiple
distinct angular directions, presumably through different macro-level
scattering or reflection paths.
Locations had up to four clusters, with an average of approximately two.
The presence of multiple clusters of paths implies
that the the possibility of both spatial multiplexing and diversity gains.

\item Applying the derived channel models to a standard cellular evaluation
framework such as \cite{3GPP36.814}, we predict that mmW systems
can offer at least an order of magnitude increase in system capacity under
reasonable assumptions on bandwidth and beamforming.
For example, we show that a hypothetical 1GHz bandwidth TDD mmW system with
a 100~m cell radii
can provide 25 times greater cell throughout than industry reported numbers for
a 20+20 MHz FDD LTE system with similar cell density.
Moreover, while the LTE capacity numbers included both single and multi-user
multi-input multi-output (MIMO),  our mmW capacity analysis did not
include any spatial multiplexing gains.
We provide strong evidence that these spatial multiplexing gains would
be significant.

\item
The system performance appears to be robust to outages provided they are at levels
similar or even a little worse than the outages we observed in the NYC
measurements.  This robustness to outage is very encouraging since
outages is one of the key concerns with mmW cellular.  However, we also show
that should outages be significantly worse than what we observed, the system
performance, particularly the cell edge rate, can be greatly impacted.
\end{itemize}

In addition to the measurement studies above,
some of the capacity analysis in this paper appeared in a conference version
\cite{AkLiuRanEr:13-GC}.  The current work provides much more extensive
modeling of the channels, more detailed discussions of the beamforming and
MIMO characteristics and simulations of features such as outage.

\subsection{Prior Measurements}

Particularly with the development of 60~GHz LAN and PAN systems, mmW signals have been
extensively characterized in indoor environments
\cite{Zwick05,Giannetti:99,Anderson04,Smulders,Manabe,ben2011millimeter,ted2}.
However, the propagation of mmW signals in outdoor settings for micro- and
picocellular networks is relatively less understood.
Due to the lack of actual measured channel data,
many earlier studies \cite{KhanPi:11,ZhangMadhow1,AkoumAyaHeath:12,PietBRPC:12}
have thus relied on either analytic models or commercial ray tracing software
with various reflection assumptions.  Below, we will compare our experimental results
with some of these models.

Also, measurements in Local Multipoint Distribution Systems at 28~GHz
-- the prior system most close to mmW cellular --
have been inconclusive:
For example, a study \cite{ElrefShak:97} found 80\% coverage at ranges
up to 1--2~km, while \cite{SeiArn:95} claimed that LOS connectivity would be
required.  Our own previous
studies at 38~GHz \cite{Rappaport:13-BBmmW,Rappaport38:12,ted:rww12,ted:wcnc12,Rappaport-72GHz:13} found that relatively long-range
links ($> 300$ m) could be established.  However, these measurements
were performed in an outdoor
campus setting with much lower building density and
greater opportunities for LOS connectivity than would be found in a typical
urban deployment.

\section{Measurement Methodology} \label{sec:chanMeas}

To assess of mmW propagation in urban environments,
our team conducted extensive measurements of 28 and 73~GHz channels in New York City.
Details of the measurements  can be found in
\cite{Rappaport:12-28G,Rappaport:28NYCPenetrationLoss,Samimi:AoAD}.
Both the 28 and 73~GHz are natural candidates for early mmW deployments.
The 28~GHz bands were previously targeted for Local Multipoint Distribution
Systems (LMDS) systems and are now an attractive
opportunity for initial deployments of mmW cellular given
their relatively lower frequency within the mmW range.
The E-Band (71-76 GHz and 81-86 GHz)~\cite{wells:09}
has abundant spectrum and adaptable for dense deployment, and could accommodate further
expansion should the lower frequencies become crowded.

\begin{figure}
\begin{center}
    \includegraphics[width=2.5in]{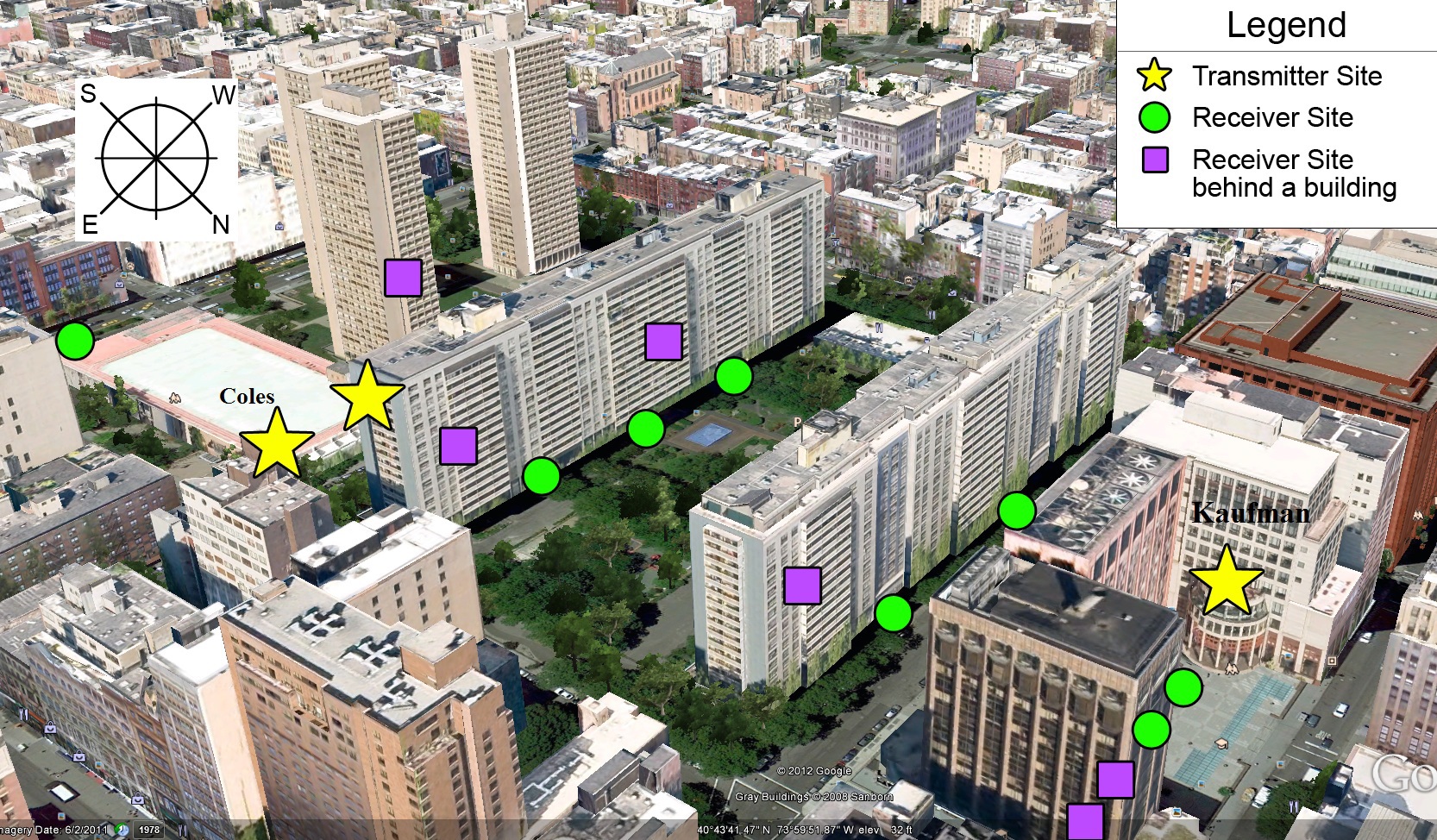}
\end{center}
\caption{Image from \cite{Rappaport:12-28G}
  showing typical measurement locations in NYC at 28~GHz.
  Similar locations were used for 73~GHz.
}
\label{fig:measMap}
\end{figure}

To measure the channel characteristics in these frequencies,
we emulated microcellular type deployments where
transmitters were placed on rooftops
7 and 17 meters (approximately 2 to 5 stories) high
and measurements were then made at a
number of street level locations up to 500~m from the transmitters
(see Fig.~\ref{fig:measMap}).
To characterize both the bulk path loss and spatial structure of the
channels, measurements
were performed with highly directional horn antennas (30 dBm RF power,
24.5~dBi gain at both TX and RX sides, and
$\approx 10^\circ$ beamwidths in both the vertical and horizontal planes
provided by rotatable horn antennas).

Since transmissions were always made from the rooftop location to the street,
in all the reported measurements below, characteristics of the transmitter
will be representative of the base station (BS) and characteristics of the receiver
will be representative of a mobile, or user equipment (UE).
At each transmitter (TX) - receiver (RX) location pair,
the azimuth (horizontal) and elevation (vertical) angles of both the transmitter and receiver were
swept to first find the direction of the maximal receive power.
After this point, power measurements were then made at various angular
offsets from the strongest angular locations.  In particular,
the horizontal angles at both the TX and RX were swept in 10$^\circ$
steps from 0 to 360$^\circ$.  Vertical angles were also sampled, typically within
a $\pm 20^\circ$ range from the horizon in the vertical plane.
At each angular sampling point, the channel sounder was used to detect
any signal paths.  To reject noise, only paths that exceeded
a 5~dB SNR threshold were included in the power-delay
profile (PDP).  Since the channel sounder has a processing gain of 30~dB,
only extremely weak paths would not be detected in this system -- See
\cite{Rappaport:12-28G,Rappaport:28NYCPenetrationLoss,Samimi:AoAD} for more details.
The power at each angular location is the sum of received powers across
all delays (i.e.\ the sum of the PDP).  A location would be considered in \emph{outage}
if there were no detected paths across all angular measurements.

\section{Channel Modeling and Parameter Estimation}

\subsection{Distance-Based Path Loss} \label{sec:distPL}
We first estimated the total omnidirectional path loss as a function of the TX-RX
distance.  At each location that was not in outage, the path loss was estimated as
\beq \label{eq:plgain}
    PL = P_{TX} - P_{RX} + G_{TX} + G_{RX},
\eeq
where $P_{TX}$ is the total transmit power in dBm,
$P_{RX}$ is the total integrated receive power over all the angular directions
and $G_{TX}$ and $G_{RX}$ are the gains
of the horn antennas.  For this experiment, $P_{TX}=30$~dBm and
$G_{TX}=G_{RX}=$~24.5 dBi.  Note that the path loss \eqref{eq:plgain}
represents an \emph{isotropic} (omnidirectional, unity antenna gain) value
i.e., the difference between the average transmit and
receive power seen in a random transmit and receive direction.
The path loss thus does not include any beamforming gains obtained
by directing the transmitter or receiver correctly
-- we will discuss the beamforming gains in detail below.

\begin{figure}
    \centering
    \includegraphics[width=3.5in]{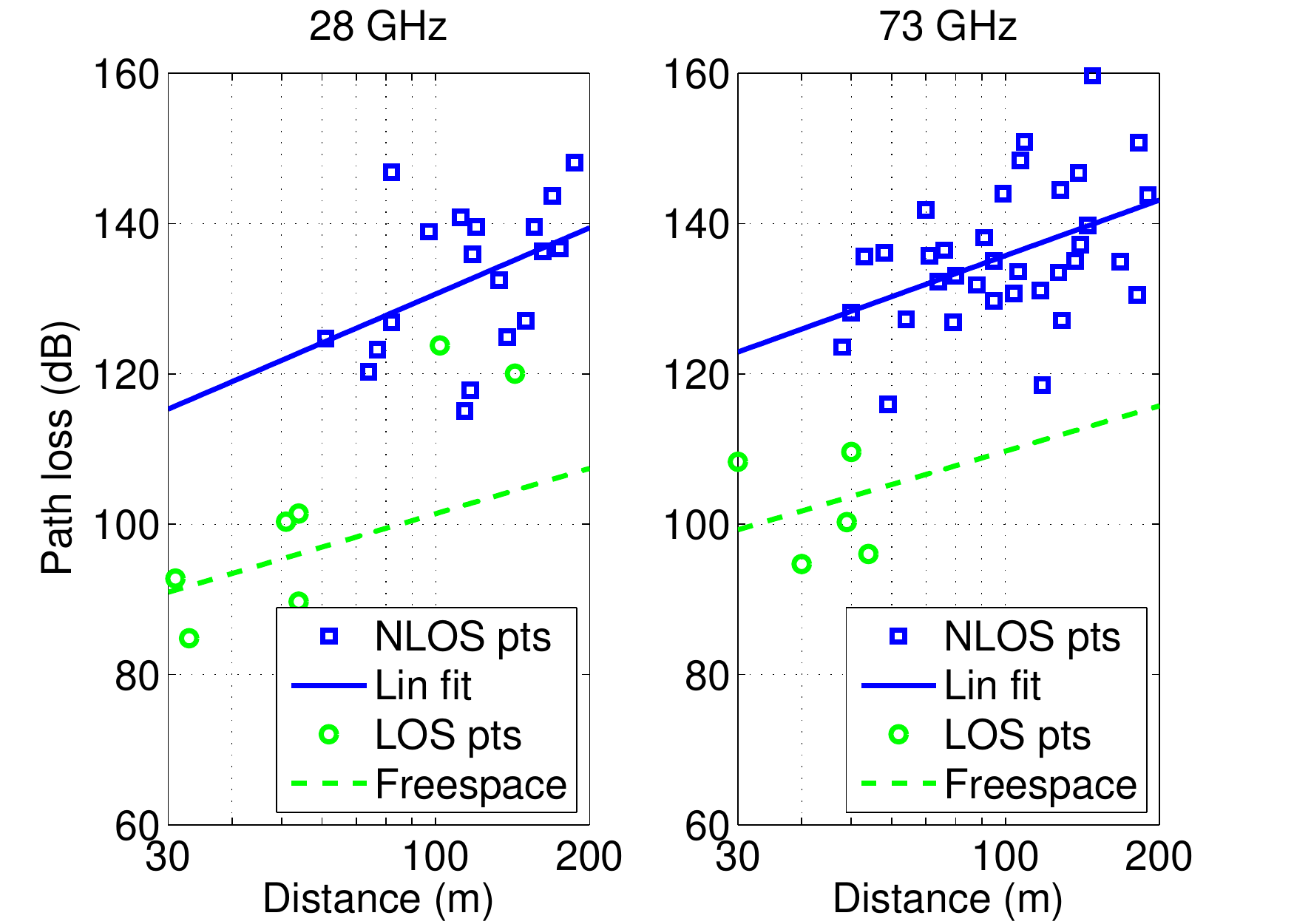}
    \caption{Scatter plot along with a linear fit of the estimated omnidirectional
     path losses as a function of the TX-RX separation for 28 and 73 GHz.}
    \label{fig:distPL}
\end{figure}

A scatter plot of the path losses at different locations
as a function of the TX-RX LOS distance
is plotted in Fig.~\ref{fig:distPL}.  In the measurements in Section~\ref{sec:chanMeas},
each location was
manually classified as either LOS, where the TX was visible to the RX,
or NLOS, where the TX was obstructed.
In standard cellular models such as \cite{3GPP36.814}, it is
common to fit the LOS and NLOS path losses separately.

For the NLOS points,
Fig.~\ref{fig:distPL} plots a fit using a standard linear model,
\beq \label{eq:plLin}
    PL(d)\mbox{ [dB]}  = \alpha + \beta 10\log_{10}(d) + \xi,
    \quad \xi \sim {\mathcal N}(0,\sigma^2),
\eeq
where $d$ is the distance in meters, $\alpha$ and $\beta$ are the least square
fits of floating intercept and slope
over the measured distances (30 to 200~m),
and $\sigma^2$ is the lognormal shadowing variance.
The values of $\alpha$, $\beta$ and $\sigma^2$ are
shown in Table~\ref{tbl:largeScaleParam}.
To assess the accuracy of the parameter estimates,
a standard Cram{\'e}r-Rao calculation shows that the standard deviation
in the median path loss due to noise was $< 2$ dB over the range of tested distances.

Note that for $f_c = $ 73~GHz,
there were two mobile antenna heights in the experiments:
4.02~m (a typical backhaul receiver height)
and 2.0~m (a typical model height).  The table provides numbers for both a mixture
of heights and for the mobile only height.  Unless otherwise stated,
we will use the mobile only height in all subsequent analysis.

For the LOS points, Fig.~\ref{fig:distPL} shows that the
theoretical free space path loss
from Friis' Law \cite{Rappaport:02} provides a good fit for
the LOS points below 100m.
However, at 28~GHz, there
are two LOS points at distances greater than 100m where the path loss is not well-fit
via a free space propagation model.  It is likely that these two points
saw higher path losses, since although the the TX was visible to the RX,
the main path arrived in a NLOS direction.
The values for $\alpha$ and $\beta$ predicted by Friis' law
and the mean-squared error $\sigma^2$ of the observed data from
Friis' Law are shown in Table~\ref{tbl:largeScaleParam}.

We should note that these numbers differ somewhat with the values reported in earlier work
\cite{Rappaport:12-28G,Rappaport:28NYCPenetrationLoss,Samimi:AoAD}.
Those works fit the path loss to power measurements for small angular regions.  Here, we
are fitting the total power over all directions.
Also, note that a close-in free space reference path loss model with a fixed leverage
point may also be used.  Such a fit is  equivalent to using the linear model \eqref{eq:plLin}
with the additional constraint that $\alpha + \beta 10\log_{10}(d_0)$ has some fixed value
for some given reference free space distance $d_0$.
Work in \cite{Rappaport-72GHz:13} shows that since this close-in free space model
has one less free parameter, the model
is less sensitive to perturbations in data, with only a slightly greater
(e.g.\ 0.5 dB standard deviation) fitting error.  While
the analysis below will not use this
fixed leverage point model, we point this out to caution against
ascribing any physical meaning to the estimated values for $\alpha$ or
$\beta$ in \eqref{eq:plLin}, and understanding that the values are somewhat sensitive to the data and should not be used outside the tested distances.

\subsection{Spatial Cluster Detection} \label{sec:cluster}

To characterize the spatial pattern of the antenna, we follow
a standard model along the lines of
the 3GPP / ITU MIMO specification \cite{3GPP36.814,ITU-M.2134}.
In the 3GPP / ITU MIMO model, the channel is assumed to be composed of a
random number $K$ of ``path clusters", each cluster corresponding to a
macro-level scattering path.
Each path cluster is described by:
\begin{itemize}
\item A fraction of the total power;
\item Central azimuth (horizontal) and elevation (vertical)
angles of departure and arrival;
\item Angular beamspreads around those central angles; and
\item An absolute propagation time group delay of the cluster and
power delay profile around the group delay.
\end{itemize}
In this work, we develop statistical models for the cluster power fractions
and angular / spatial characteristics.  However, we do not study temporal characteristics
such as the relative propagation times or the time delay profiles.  Due to the nature
of the measurements, obtaining relative propagation times from different angular directions
requires further analysis and will be subject of a forthcoming paper.
The models here are only narrowband.

To fit the cluster model to our data,
our first step was to detect the  path clusters
in the angular domain at each TX-RX location pair.
As described above in Section~\ref{sec:chanMeas}, at each location
pair, the RX power was measured at various angular offsets.
Since there are horizontal and vertical angles at both the transmitter
and receiver, the measurements can be interpreted as a
sampling of power measurements in a four-dimensional space.

\begin{figure}
    \centering
    \includegraphics[width=3.5in]{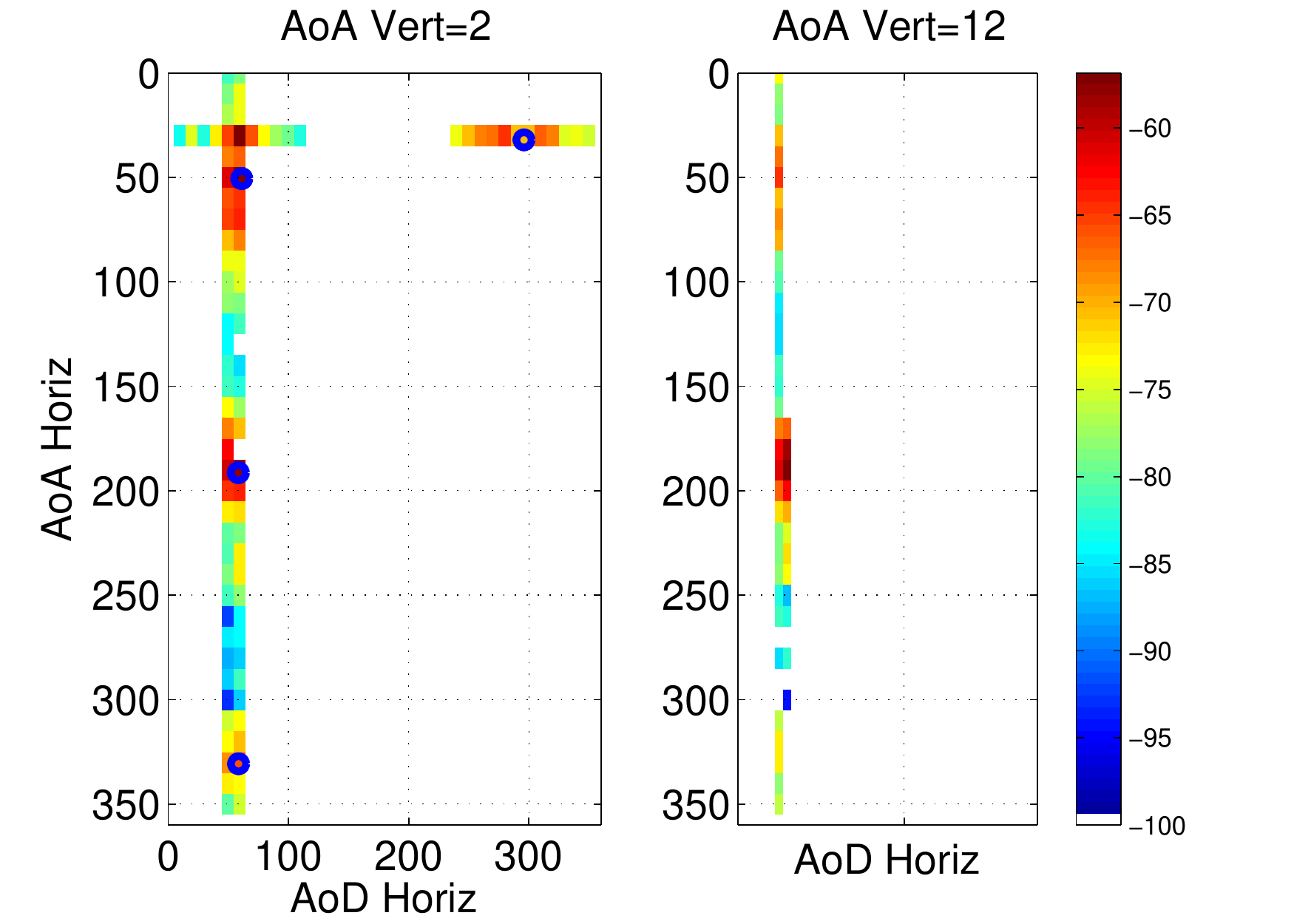}
    \caption{Typical RX power angular profile at 28 GHz.  Colors represent
    the average RX power in dBm at each angular offset, with white areas
    representing angular offsets that were either not measured, or
    had too low power to be validly detected.  The blue circles
    represent the detected path cluster centers from our
    path clustering algorithm.}
    \label{fig:rxHeatMap}
\end{figure}

A typical RX profile is shown in Fig.~\ref{fig:rxHeatMap}.
Due to time, it was impossible to measure the entire four-dimensional
angular space.  Instead, at each location,
only a subset of the angular offsets were measured.
For example, in the location depicted in Fig.~\ref{fig:rxHeatMap},
the RX power was measured along two strips:
one strip where the horizontal AoA was swept from 0 to 360 with the
horizontal (azimuth) AoD varying in a 30 degree interval; and a second strip where
the horizontal AoA was constant and the horizontal AoD was varied from
0 to 360.  Two different values for the vertical (elevation) AoA were taken --
the power measurements in each vertical AoA
shown in three different subplots in Fig.~\ref{fig:rxHeatMap}.
The vertical AoD was
kept constant since there was less angular dispersion in that dimension.
This measurement pattern was fairly typical, although in the 73~GHz measurements,
we tended to measure more vertical AoA points.

The locations in white in Fig.~\ref{fig:rxHeatMap} represent angular
points where either the power was not measured, or the insufficient
signal power was detected.  Sufficient receive power to be validly
was defined as finding at least
a single path with 5dB~SNR above the thermal noise.  The power
in all points that were either not measured or insufficient power
was detected was treated as zero.  If no valid
angular points were detected, the location was considered in outage.

Detection of the spatial clusters amounts to finding regions in the
four-dimensional angular space where the received energy is concentrated.
This is a classic clustering problem, and for each candidate number of
clusters $K$, we used a
standard $K$-means clustering algorithm \cite{Bishop:06}
to approximately find $K$ clusters in the receive power domain with minimal
angular dispersion.
The $K$-means algorithm groups all the validly detected angular points into one of $K$
clusters.  For channel modeling in this paper, we use the algorithm to identify
clusters with minimal angular variance as weighted by the receive power.
The $K$-means algorithm performs this clustering by alternately (i) identifying the
power weighted centroid of each cluster given a classification of the angular points
into clusters; and (ii) updating the
cluster identification by associating each angular point with its closest cluster center.

The clustering algorithm was run with increasing values of $K$, stopping when either of the
following conditions were
satisfied:  (i) any two of the $K$ detected clusters were within
2 standard deviations in all angular directions; or
(ii) one of the clusters were empty.
In this way, we obtain at each location, an estimate of the number of resolvable clusters $K$,
their central angles, root-mean-squared angular spreads, and receive power.
In the example location in Fig.~\ref{fig:rxHeatMap}, there were four detected clusters.
The centers are shown in the left plot in the blue circles.

\subsection{Cluster Parameters}

After detecting the clusters and the corresponding cluster parameters,
we fit the following statistical models to the various cluster features.

\begin{figure}
    \centering
    \includegraphics[width=3.5in]{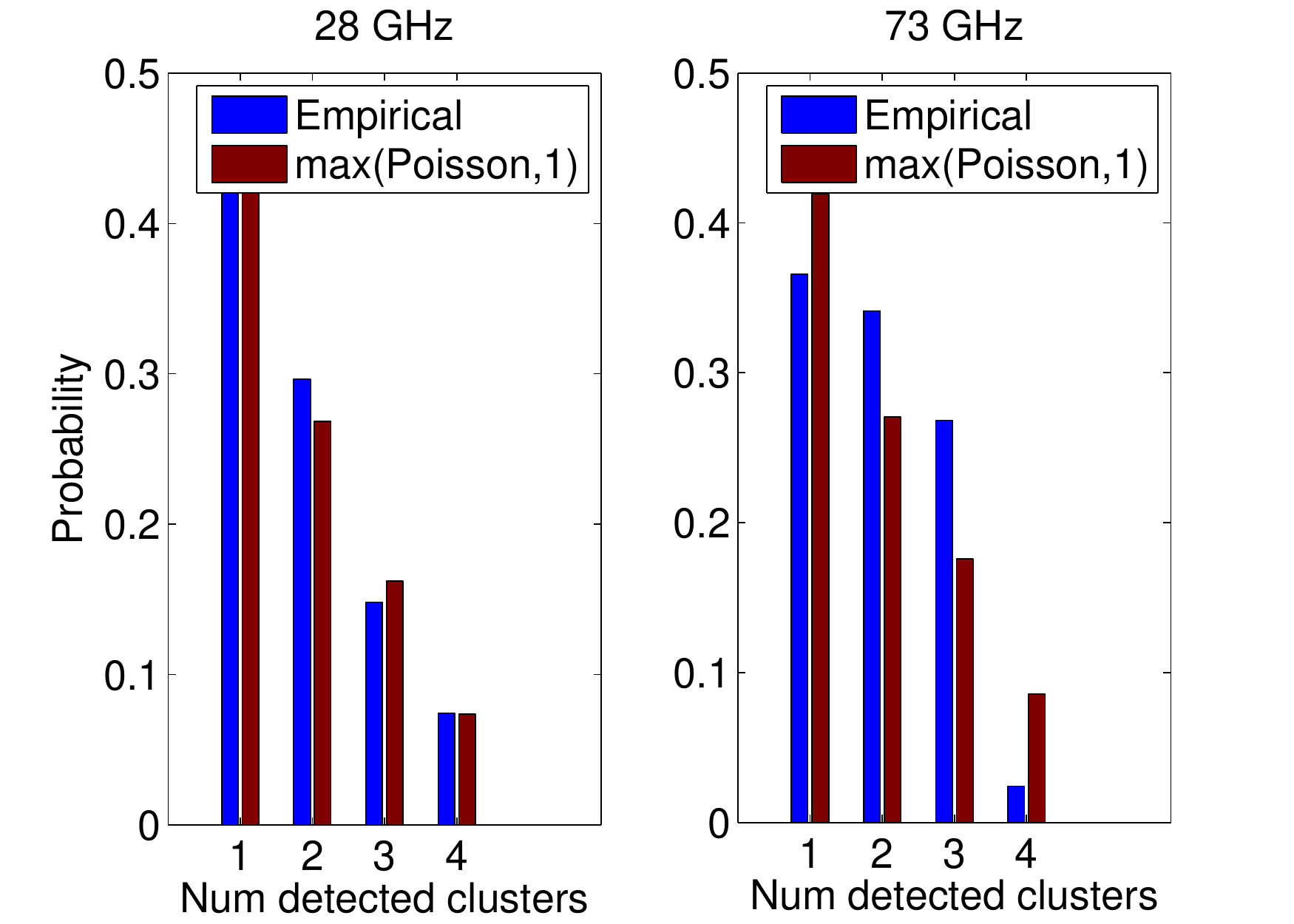}
    \caption{Distribution of the number of detected clusters
    at 28 and 73~GHz.
    The measured distribution is labeled 'Empirical', which
    matches a Poisson distribution \eqref{eq:KPoisson} well. }
    \label{fig:KPoisson}
\end{figure}

\paragraph{Number of clusters}
At the locations where a signal was detected (i.e.\ not in outage),
the number of estimated clusters
detected by our clustering algorithm, varied from 1 to 4.  The measured
distribution is plotted in the bar graph in Fig.~\ref{fig:KPoisson}
in the bars labeled ``empirical".  Also, plotted is the distribution for a
random variable $K$ of the form,
\beq \label{eq:KPoisson}
    K \sim \max\{\mathop{Poisson}(\lambda), 1 \},
\eeq
where $\lambda$ set to empirical mean of $K$.
It can be seen that this Poisson-max distribution is a good fit
to the true number of detected clusters, particularly for 28 GHz.

\paragraph{Cluster Power Fraction}
A critical component in the model is the distribution of power amongst the clusters.
In the 3GPP model \cite[Section B.1.2.2.1]{3GPP36.814}, the cluster power fractions
are modeled as follows:  T
First, each cluster $k$ has an absolute group delay, $\tau_k$,
that is assumed to be exponentially distributed.  Therefore, we can write $\tau_k$
as
\beq \label{eq:tauExp}
    \tau_k = -r_\tau\sigma_\tau \log U_k
\eeq
for a uniform random variable $U_k \sim U[0,1]$ and constants $r_\tau$
and $\sigma_\tau$.
The cluster $k$ is assumed to have a power that scales by
\beq \label{eq:powFrac0}
    \gamma_k' = \exp\left[ \tau_k\frac{r_\tau-1}{\sigma_\tau r_\tau}\right]
        10^{-0.1Z_k}, \quad Z_k\sim {\mathcal N}(0,\zeta^2),
\eeq
where  the first term in the product
places an exponential decay in the cluster power with the
delay $\tau_k$, and the second term accounts for lognormal variations in the
per cluster power with some variance $\zeta^2$.
The final power fractions for the different clusters are then found by
normalizing the values in \eqref{eq:powFrac0} to unity, so that the
fraction of power in $k$-th cluster is given by
\beq \label{eq:powFrac1}
    \gamma_k = \frac{\gamma_k'}{\sum_{j=1}^K \gamma_j'}.
\eeq

In the measurements in this study, we do not know the relative propagation
delays $\tau_k$ of the different clusters, so we treat them as unknown
latent variables.  Substituting \eqref{eq:tauExp} into \eqref{eq:powFrac0},
we obtain
\beq \label{eq:powFrac2}
    \gamma_k' = U_k^{r_\tau-1} 10^{-0.1Z_k},
        \quad U_k\sim U[0,1],
        \quad Z_k \sim {\mathcal N}(0,\zeta^2),
\eeq
The constants $r_\tau$ and $\zeta^2$ can then be treated as model parameters.
Note that the lognormal variations $Z_k$ in the per cluster
power fractions \eqref{eq:powFrac2} are distinct from the lognormal variations in
total omnidirectional path loss \eqref{eq:plLin}.

\begin{figure}
    \centering
    \includegraphics[width=3.5in]{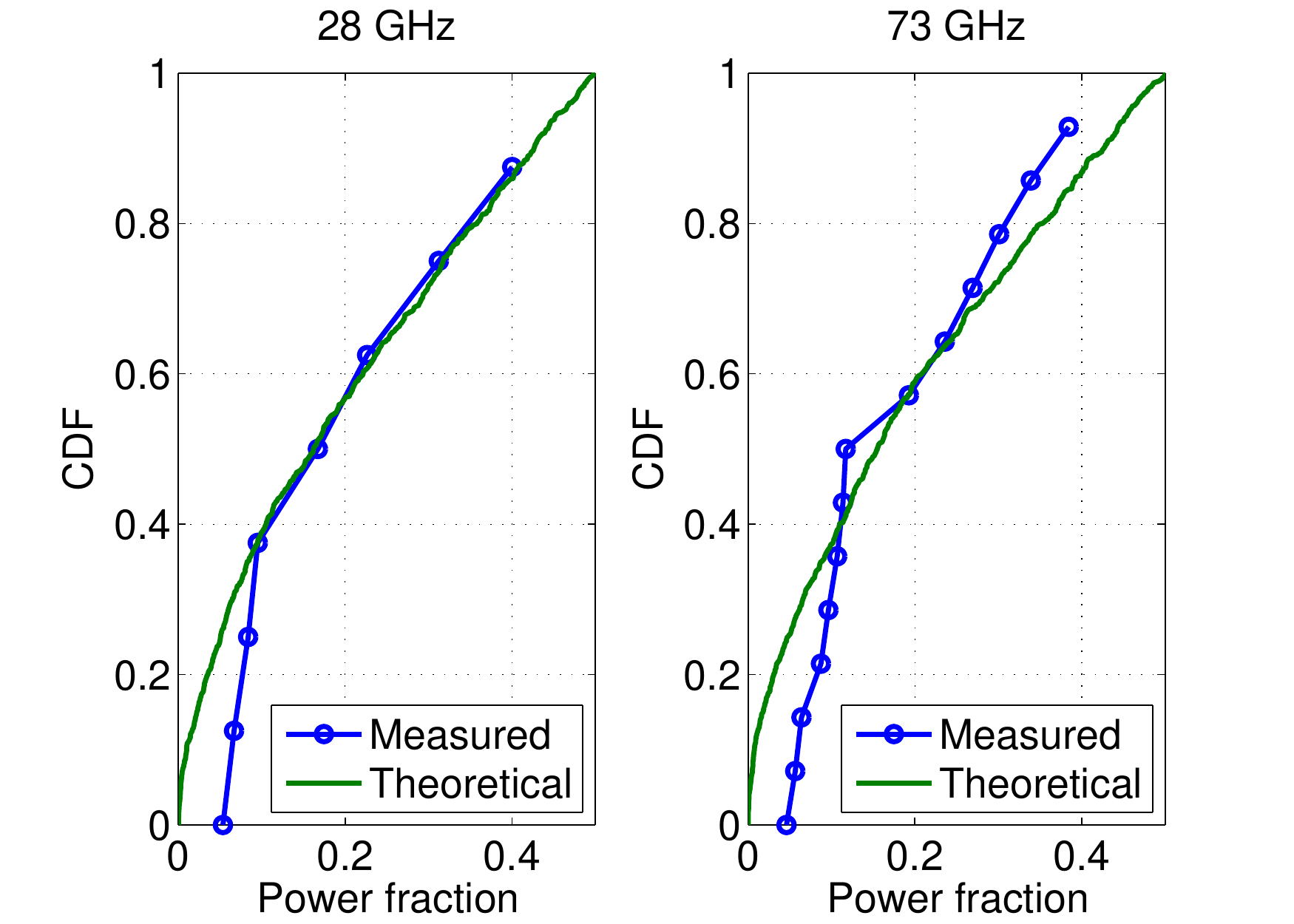}
    \caption{Distribution of the fraction of power in the
     weaker cluster, when $K=2$ clusters were detected.
     Plotted are the measured distributions and the best fit of
     the theoretical model in \eqref{eq:powFrac1} and \eqref{eq:powFrac1}. }
    \label{fig:clusPowFrac}
\end{figure}

For the mmW data, Fig.~\ref{fig:clusPowFrac} shows the distribution of
the fraction of power in the weaker cluster in the case when $K=2$ clusters were
detected.   Also plotted is the theoretical distribution based on \eqref{eq:powFrac1}
and \eqref{eq:powFrac2} where the parameters $r_\tau$ and $\zeta^2$ were
fit via an approximate maximum likelihood method.
Since the measurement data we have does not have the relative delays
of the different clusters we treat the variable $U_k$ in \eqref{eq:powFrac1}
as an unknown latent variable, adding to the variation in the cluster power
distributions.
The estimated ML parameters are shown in Table~\ref{tbl:largeScaleParam},
with the values in 28 and 73~GHz being very similar.

We see that the 3GPP model with the ML parameter selection provides
an excellent fit for the observed power fraction for clusters with more than
10\% of the energy. The model is likely not fitting the very low energy
clusters since our cluster detection is likely unable to find those clusters.
However, for cases where the clusters have significant power, the model
appears accurate.
Also, since there were very few locations
where the number of clusters was $K \geq 3$, we only fit the parameters
based on the $K=2$ case.  In the simulations below, we will assume the model
is valid for all $K$.

\begin{figure}
    \centering
    \includegraphics[width=3.5in]{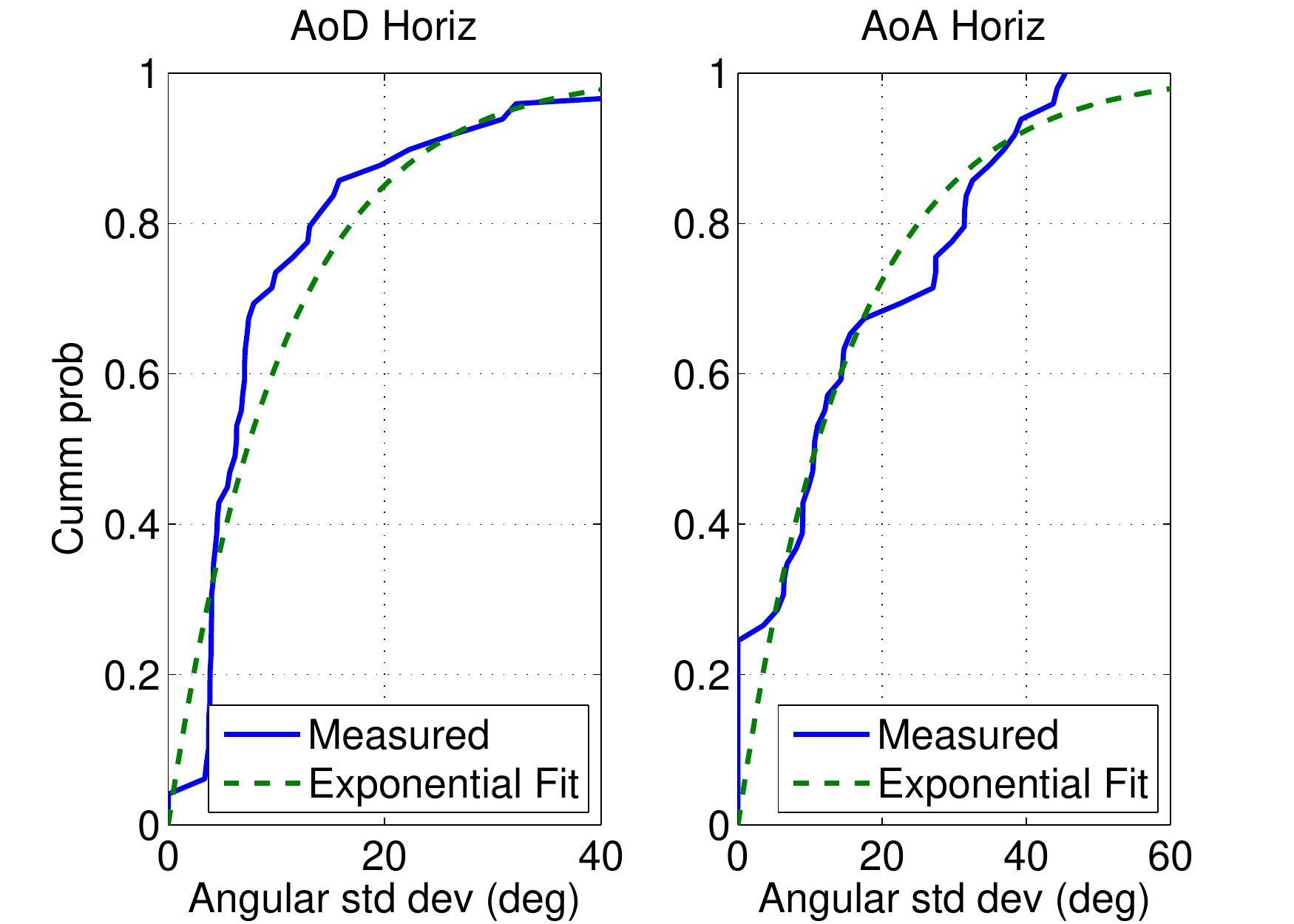}
    \caption{Distribution of the rms angular spreads
    in the horizontal (azimuth) AoA and AoDs. Also plotted is an
    exponential distribution with the same empirical mean. }
    \label{fig:angDisp}
\end{figure}

\paragraph{Angular Dispersion}
For each detected cluster, we measured the root mean-squared (rms)
beamspread in the different angular dimensions.  In
the angular spread estimation
in each cluster, we excluded power measurements from the lowest
10\% of the total cluster power.
This clipping introduces a small bias in the angular spread estimate.
Although these low power points correspond to valid
signals (as described above, all power measurements were only
admitted into the data set if the signals were received
with a minimum power level),
the clipping reduced the sensitivity to misclassifications of points
at the cluster boundaries.
The distribution of the angular spreads at 28~GHz computed in this manner
is shown in Fig.~\ref{fig:angDisp}.
Based on \cite{winner2:07}, we have also plotted an
exponential distribution with the same empirical mean.
We see that the exponential distribution provides a
good fit of the data.  Similar distributions were observed at 73~GHz, although
they are not plotted here.

\subsection{LOS, NLOS, and Outage Probabilities}  \label{sec:pblk}

Up to now, all the model parameters were based on locations not in outage.
That is, there was some power detected in at least one delay in one angular location
-- See Section~\ref{sec:chanMeas}.
However, in many locations, particularly locations $>$ 200m from the transmitter,
it was simply impossible to detect any signal with
transmit powers between 15 and 30 dBm.
This outage is likely due to environmental obstructions that occlude all paths
(either via reflections or scattering) to the receiver. The presence of outage in this manner is perhaps the most significant difference moving from conventional microwave / UHF to millimeter wave frequencies, and requires accurate modeling to properly assess system performance.

Current 3GPP evaluation methodologies
such as \cite{3GPP36.814} generally use a statistical model
where each link is in either a LOS or NLOS state, with the probability of being
in either state being some function of the distance.  The path loss and other
link characteristics are then a function of the link state, with potentially
different models in the LOS and NLOS conditions.
Outage occurs implicitly when the path loss in either the LOS or NLOS state
is sufficiently large.

For mmW systems, we propose to add an additional state, so that each link
can be in one of three conditions:  LOS, NLOS or outage.  In the outage
condition, we assume there is no link between the TX and RX --- that is, the path loss
is infinite.  By adding this third state with a random probability
for a complete loss, the model provides
a better reflection of outage possibilities inherent in mmW.
As a statistical model, we assume probability functions for the three states
are of the form:
\begin{subequations} \label{eq:phybrid}
\beqa
     p_{\outage}(d) &=& \max(0,1-e^{-a_{\out}d+b_{\out}}) \label{eq:Pout} \\
     p_{\LOS}(d) &=&  (1-p_{\outage}(d))e^{-a_{\los}d} \label{eq:Plos} \\
     p_{\NLOS}(d) &=&  1-p_{\outage}(d) - p_{\LOS}(d) \label{eq:Pnlos}
\eeqa
\end{subequations}
where the parameters $a_{los}$, $a_{out}$ and $b_{out}$ are parameters
that are fit from the data.
The outage probability model \eqref{eq:Pout} is similar in form to the
3GPP suburban relay-UE NLOS model \cite{3GPP36.814}.
The form for the LOS probability \eqref{eq:Plos} can be derived on the basis
of random shape theory \cite{bai2013analysis}.
See also \cite{Heath:14arXiv} for a discussion on the outage modeling and its effect
on capacity.

\begin{figure}
	\includegraphics[width=3.5in]{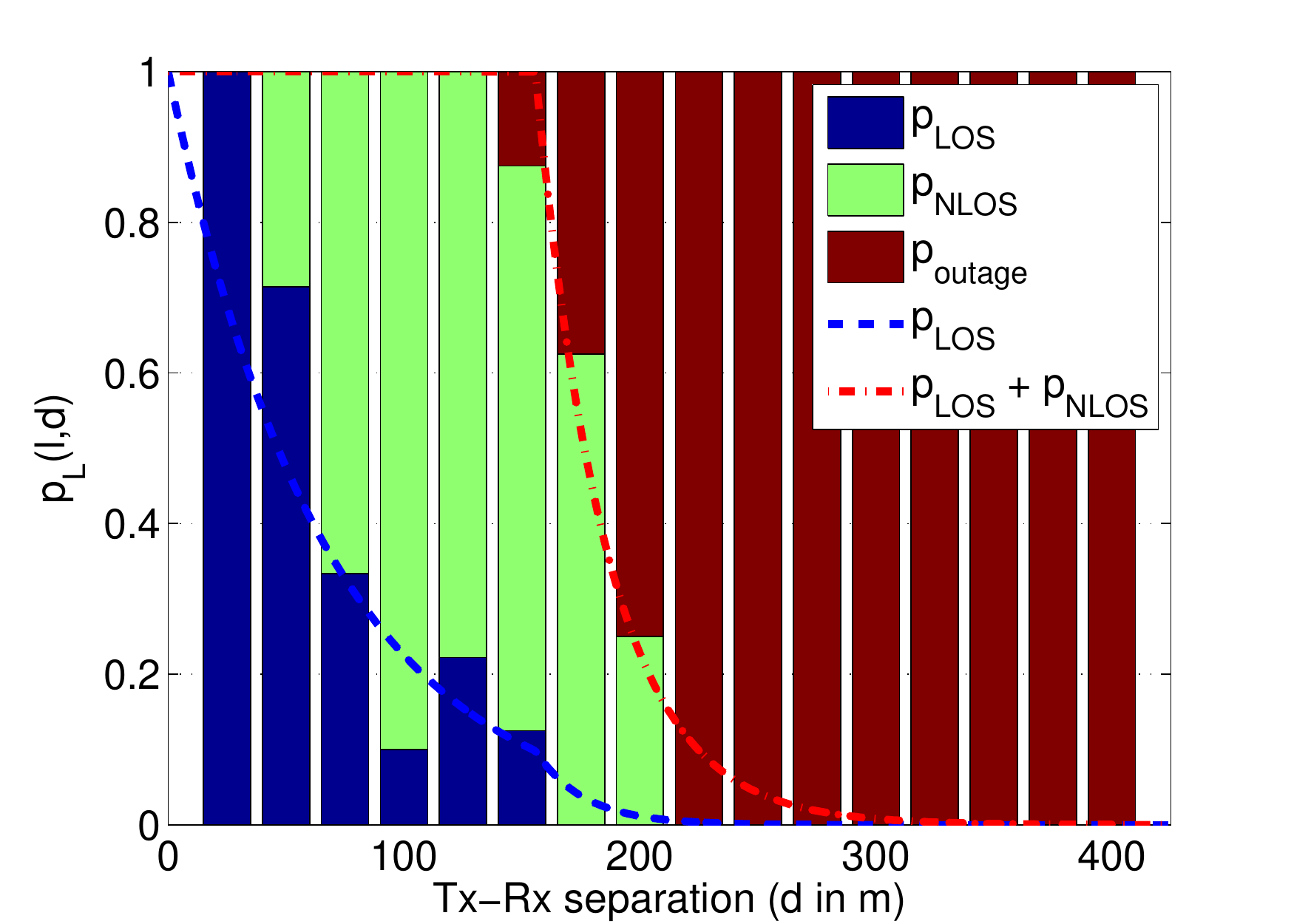}
	\caption{The fitted curves and the empirical values of $p_{\LOS}(d)$, $p_{\NLOS}(d)$, and $p_{\outage}(d)$ as a function of the distance $d$.
Measurement data is based on 42 TX-RX location pairs
with distances from 30~m to 420~m at 28~GHz.}
	\label{fig:blkProb}
\end{figure}

The parameters in the models were fit based on maximum likelihood
estimation from the 42 TX-RX location pairs in the 28~GHz measurements in
\cite{Sun-Beam:13,Nie28-73SOS:14}.  In the simulations below,
we assumed that the same probabilities
held for the 73~GHz.  The values are shown in Table~\ref{tbl:largeScaleParam}.
Fig.~\ref{fig:blkProb} shows the fractions of points that were observed
to be in each of the three states -- outage, NLOS and LOS.  Also plotted is the
probability functions in \eqref{eq:phybrid} with the ML estimated parameter
values.  It can be seen that the probabilities provide an excellent fit.

That being said, caution should be exercised in generalizing these particular
parameter values to other scenarios.  Outage conditions
are highly environmentally dependent,
and further study is likely needed to find parameters that are valid across a range
of circumstances.  Nonetheless, we believe that the experiments illustrate that
a three state model with an explicit outage state can provide an better description
for variability in mmW link conditions.  Below we will see assess the sensitivity
of the model parameters to the link state assumptions.

\subsection{Small-Scale Fading Simulation} \label{sec:smallScale}

The statistical models and  parameters are summarized in
Table~\ref{tbl:largeScaleParam}.  These parameters all represent
large-scale fading characteristics, meaning they are parameters associated
with the macro-scattering environment and change
relatively slowly \cite{Rappaport:02}.

\begin{table*}
\caption{Proposed Statistical Model for the Large-scale Parameters
 based on the NYC data in  \cite{rappaportmillimeter}. }
  \label{tbl:largeScaleParam}
\begin{threeparttable}

  \begin{tabular}{|>{\raggedright}p{1.4in}|>{\raggedright}p{1.8in}|
    >{\raggedright}p{1.55in}|>{\raggedright}p{1.75in}|}
	\hline
	\textbf{Variable}  &  \textbf{Model} &
    \multicolumn{2}{c|}{\textbf{Model Parameter Values} }
    \tabularnewline    \cline{3-4}
    & & \textbf{28 GHz} & \textbf{73 GHz}     \tabularnewline  \hline
Omnidirectional path loss, $PL$\\  and lognormal shadowing, $\xi$ & $PL = \alpha+10\beta\log_{10}(d) + \xi$ [dB]\\$\xi \sim \mathcal{N}(0,\sigma^2)$,
    $d$ in meters  & NLOS:\\ $\alpha=72.0$, $\beta=2.92$,  $\sigma = 8.7$ dB \\ \vspace{\baselineskip} LOS:\\ $\alpha=61.4$, $\beta=2$,  $\sigma = 5.8$ dB & NLOS:\\ $\alpha=86.6$, $\beta=2.45$, $\sigma=8.0$ dB ($\dagger$) \\ $\alpha=82.7$, $\beta=2.69$, $\sigma=7.7$ dB ($\ddagger$)\\ LOS:\\ $\alpha=69.8$, $\beta=2$,  $\sigma = 5.8$ dB
        \tabularnewline \hline

	NLOS-LOS-Outage probability  & See~ \eqref{eq:phybrid} &
    \multicolumn{2}{l|}{$a_\out = 0.0334\mathrm{m}^{-1}$,   $b_\out = 5.2$,    $a_\los = 0.0149\mathrm{m}^{-1}$}
    \tabularnewline \hline	

	Number of clusters, $K$ & $K \sim \max\{\mathop{Poisson}(\lambda),1\}$  &
    $\lambda=1.8$ & $\lambda=1.9$ \tabularnewline \hline	

    Cluster power fraction    &    See \eqref{eq:powFrac1} and \eqref{eq:powFrac2}:
    $\gamma_k' = U_k^{r_\tau-1} 10^{0.1Z_k}$, $Z_k\sim {\mathcal N}(0,\zeta^2)$,
    $U_k \sim U[0,1]$
    & $r_\tau=2.8$, $\zeta=4.0$ & $r_\tau=3.0$, $\zeta=4.0$
        \tabularnewline \hline
    	
    BS and UE horizontal cluster central angles, $\theta$ &
        $\theta \sim U(0,2\pi)$ & &  \tabularnewline \hline

    BS and UE vertical cluster central angles, $\phi$ &
        $\phi$ = LOS elevation angle & &  \tabularnewline \hline

	BS cluster rms angular spread & $\sigma$ is exponentially distributed,\\
    $\Exp(\sigma) = \lambda^{-1}$   &
         \pbox{1in}{ \vspace{2mm} Horiz $\lambda^{-1} = 10.2^\circ$;
         \\ Vert $\lambda^{-1}=0^\circ$ (*) }&
         \pbox{1in}{\vspace{2mm} Horiz $\lambda^{-1} = 10.5^\circ$; \\
          Vert $\lambda^{-1}=0^\circ$ (*) }
         \tabularnewline[3ex] \hline

	UE rms angular spread & $\sigma$ is exponentially distributed,\\
 $\Exp(\sigma) = \lambda^{-1}$ &
         \pbox{1in}{ \vspace{2mm} Horiz $\lambda^{-1} = 15.5^\circ$;
         \\ Vert $\lambda^{-1}=6.0^\circ$ }&
         \pbox{1in}{\vspace{2mm} Horiz $\lambda^{-1} = 15.4^\circ$; \\
          Vert $\lambda^{-1}=3.5^\circ$ }
         \tabularnewline[3ex] \hline

  \end{tabular}
  \begin{tablenotes}
  \item Note:  The model parameters are derived in
based on converting the directional measurements
from the NYC data in  \cite{rappaportmillimeter},
and assuming an isotropic (omnidirectional, unity gain) channel model
with the 49 dB of antenna gains removed from the measurements.
	\item ($\dagger$) Parameters for the 2m-RX-height data and 4.06m-RX-height data combined.
	\item ($\ddagger$) Parameters for the 2m-RX-height data only.
  \item (*) BS downtilt
was fixed at 10 degree for all measurements, resulting in no measurable vertical angular
spread at BS.
  \end{tablenotes}
  \end{threeparttable}
\end{table*}

One can generate a random narrowband time-varying
channel gain matrix for these parameters following a similar
procedure as the 3GPP / ITU model
\cite{3GPP36.814,ITU-M.2134} as follows:  First, we generate random realizations of all the large-scale
parameters in Table~\ref{tbl:largeScaleParam} including the
distance-based omni path loss, the number of clusters $K$,
their power fractions, central angles and angular beamspreads.
For the small-scale fading model, each of the $K$ path clusters
can then be synthesized with a large number, say $L=20$, of subpaths.
Each subpath will
have horizontal and vertical AoAs, $\theta^{rx}_{k\ell}$, $\phi^{rx}_{k\ell}$,
and horizontal and vertical AoDs, $\theta^{tx}_{k\ell}$, $\phi^{tx}_{k\ell}$,
where $k=1,\ldots,K$ is the cluster index and $\ell=1,\ldots,L$ is the
subpath index within the cluster.
These angles can be generated as wrapped Gausians around the cluster central angles
with standard deviation given by the rms angular spreads for the cluster.
Then, if there are $n_{rx}$ RX antennas and $n_{tx}$ TX antennas,
the narrowband time-varying channel gain between a TX-RX pair
can be represented by a matrix (see, for example, \cite{TseV:07} for more details):
\beq \label{eq:Hsum}
    \Hbf(t) = \frac{1}{\sqrt{L}} \sum_{k=1}^K \sum_{\ell = 1}^L
        g_{k\ell}(t) \ubf_{rx}(\theta^{rx}_{k\ell},\phi^{tx}_{k\ell})
        \ubf_{tx}^*(\theta^{tx}_{k\ell},\phi^{tx}_{k\ell}),
\eeq
where $g_{k\ell}(t)$ is the complex small-scale fading gain on the $\ell$-th
subpath of the $k$-th
cluster and $\ubf_{rx}(\cdot) \in \C^{n_{rx}}$ and $\ubf_{tx}(\cdot) \in \C^{n_{tx}}$
are the vector response functions for the RX and TX antenna arrays to the angular arrivals and
departures.  The small-scale coefficients would be given by
\[
    g_{k\ell}(t) = \gbar_{k\ell}e^{2\pi itf_{dmax}\cos(\omega_{k\ell})}, \quad
    \gbar_{k\ell} \sim {\mathcal CN}(0,\gamma_k 10^{-0.1PL}),
\]
where $f_{dmax}$ is the maximum Doppler shift,
$\omega_{k\ell}$ is the angle of arrival of the subpath relative to
the direction of motion and $PL$ is the omnidirectional path loss.
The relation between $\omega_{k\ell}$ and the angular arrivals
$\theta^{rx}_{k\ell}$ and $\phi^{rx}_{k\ell}$ will depend on the orientation of the mobile
RX array relative to the motion.
Note that the model \eqref{eq:Hsum} is only a narrowband model
since we have not yet characterized the delay spread.

\section{Comparison to 3GPP Cellular Models}

\subsection{Path Loss Comparison} \label{sec:plComp}

\begin{figure}
    \centering
    \includegraphics[width=3in]{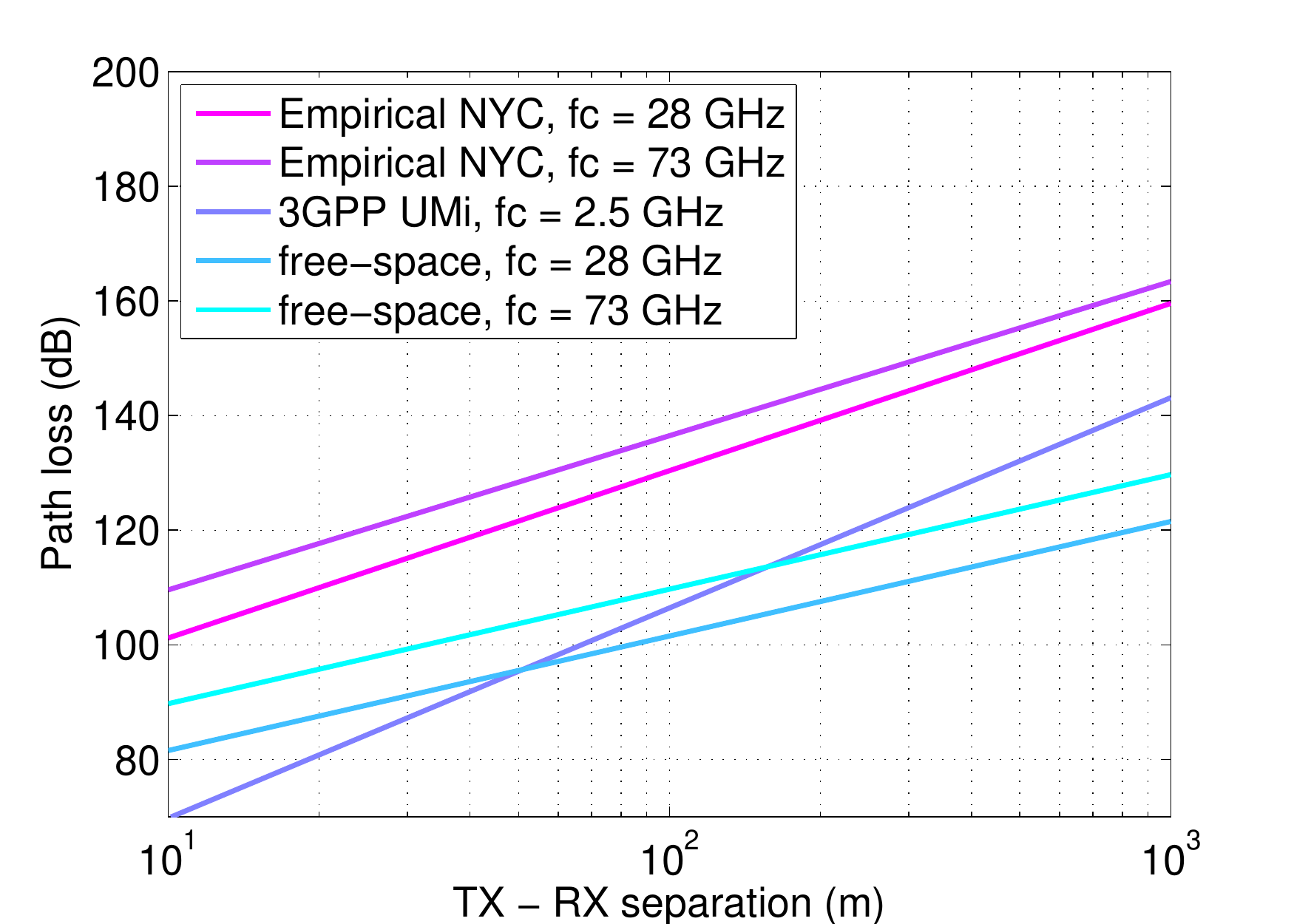}
    \caption{Comparison of distance-based path loss models.
    The curves labeled ``Empirical NYC"
    are the mmW models derived in this paper for 28 and 73~GHz.
    These are compared to
    free-space propagation for the same frequencies
    and 3GPP Urban Micro (UMi) model for 2.5~GHz. }
    \label{fig:plComp}
\end{figure}

It is useful to briefly compare the distance-based path loss we observed
for mmW signals with models for conventional cellular systems.
To this end, Fig.~\ref{fig:plComp} plots the median effective
total path loss as a function of distance for several different models:
\begin{itemize}
\item \emph{Empirical NYC:}  These curves are the omnidirectional
path loss predicted by our linear model \eqref{eq:plLin}.  Plotted is the median
path loss
\beq \label{eq:PLmed}
    PL(d) \mbox{ [dB]} = \alpha + 10\beta\log_{10}(d),
\eeq
where $d$ is the distance and the $\alpha$ and $\beta$ parameters
are the NLOS values in Table~\ref{tbl:largeScaleParam}.  For 73~GHz,
we have plotted the 2.0 UE height values.

\item \emph{Free space:}  The theoretical free space path loss is
given by Friis' Law~\cite{Rappaport:02}.
We see that, at $d=$ 100~m, the
free space path loss is approximately 30~dB
less than the model we have experimentally measured here.
Thus, many of the works such as
\cite{ZhangMadhow1,PietBRPC:12}
that assume free-space propagation may be somewhat optimistic in
their capacity predictions.
Also, it is interesting to point out that
one of the models assumed in the Samsung study \cite{KhanPi:11}
(PLF1) is precisely free space propagation + 20 dB -- a correction factor
that is also somewhat more optimistic than our experimental findings.

\item \emph{3GPP UMi:}  The standard 3GPP urban micro (UMi) path loss
model with
hexagonal deployments~\cite{3GPP36.814} is given by
\beq \label{eq:UMiPL}
    PL(d) \mbox{ [dB]}= 22.7 + 36.7\log_{10}(d) + 26 \log_{10}(f_c),
\eeq
where $d$ is distance in meters and $f_c$ is the carrier frequency in GHz.
Fig.~\ref{fig:plComp} plots this path loss model at $f_c=$ 2.5 GHz.
We see that our propagation models at both 28 and 73 GHz predict
omnidirectional path losses that, for most of the distances,
are approximately 20 to 25~dB higher than the 3GPP UMi model at 2.5~GHz.
However, since the wavelengths at 28 and 73~GHz are approximately 10 to 30
times smaller, this path loss can be entirely compensated
with sufficient beamforming on either the transmitter or
receiver with the same physical antenna size.
Moreover, if beamforming is applied on both ends, the effective path loss
can be even lower in the mmW range.
We conclude that, barring outage events and maintaining the same
physical antenna size, \emph{mmW signals do not imply any reduction
in path loss relative to current cellular frequencies, and in fact,
be improved over today's systems}.
\end{itemize}

\subsection{Spatial Characteristics}
We next compare the spatial characteristics of the mmW and microwave models.
To this end, we can compare the
experimentally derived mmW parameters in
Table~\ref{tbl:largeScaleParam} with those, for example, in
\cite[Table B.1.2.2.1-4]{3GPP36.814} for the 3GPP urban microcell model
-- the layout that would be closest to our deployment.
We immediately see that the angular spread of the clusters are similar
in the mmW and 3GPP UMi models.
While the 3GPP UMi model has somewhat more clusters,
it is possible that multiple distinct clusters were present in the mmW
scenario, but were not visible since we did not perform any temporal
analysis of the data.  That is, in our clustering algorithm above,
we group power from different time delays together in each angular offset.

Another interesting comparison is the delay scaling parameter, $r_\tau$,
which governs how relative propagation delays between clusters affects their
power faction.  Table~\ref{tbl:largeScaleParam} shows values of $r_\tau$ of 2.8 and 3.0,
which are in the same range as the values in the 3GPP UMi model
\cite[Table B.1.2.2.1-4]{3GPP36.814} suggesting that the power delay may be similar.
This property would, however, require further confirmation with
actual relative propagation delays between clusters.

\subsection{Outage Probability}

One final difference that should be noted is the outage probability.
In the standard 3GPP models, the event that a channel is completed
obstructed is not explicitly modeled.  Instead, channel variations are
accounted for by lognormal shadowing along with, in certain models,
wall and other obstruction losses.
However, we see in our experimental measurements that channels in the mmW range
can experience much more significant blockages that are not well-modeled via these
more gradual terms.  We will quantify the effects of the outages
on the system capacity below.

\section{Channel Spatial Characteristics and MIMO Gains}

A significant gain for mmW systems
derives from the capability of high-dimensional beamforming.
Current technology can easily support antenna arrays with 32 elements and higher
\cite{Doan:04,Doan:05,ZhaLiu:09,gutierrez2009chip,Nsenga:10,Ted:60Gstate11,Rajagopal:mmWMobile,Huang:2008:MWA:1524107,Rusek:13}.
Although our simulations below will assess the precise beamforming gains
in a micro-cellular type deployment, it is useful to first
consider some simple spatial statistics of the channel to qualitatively
understand how large the beamforming gains may be and how they can be practically
achieved.

\subsection{Beamforming in Millimeter Wave Frequencies}
However, before examining the channel statistics, we need to point
out two unique aspects of beamforming and spatial multiplexing in
the mmW range.
First, a full digital front-end
with high resolution A/D converters on each antenna across the wide bandwidths
of mmW systems may be prohibitive in terms of cost and power, particularly for mobile devices
\cite{Ted:60Gstate11,KhanPi:11-CommMag,KhanPi:11,Heath:partialBF}.
Most commercial designs have thus assumed phased-array architectures where signals
are combined either in RF with phase shifters \cite{ParkZim:02,KohReb:07,KohReb:09}
or at  IF \cite{Crane-Patent:88,RamBaRe:98,GuanHaHa:04} prior to the A/D conversion.
While greatly reducing the front-end power consumption,
this architecture may limit the number of separate spatial streams
that can be processed since
each spatial stream will require a separate phased-array and associated RF chain.
Such limitations will be particularly important at the UE.

A second issue is the channel coherence:  due to the high Doppler frequency it may not be feasible
to maintain the channel state information (CSI) at the transmitter, even in TDD.
In addition, full CSI at the receiver may also not be available since the beamforming must be
applied in analog and hence the beam may need to be selected without separate digital
measurements on the channels on different antennas.

\subsection{Instantaneous vs.\ Long-Term Beamforming} \label{sec:bfGain}

Under the above constraints, we begin by trying to assessing what the rough
gains we can expect from beamforing are as follows:
Suppose that the transmitter and receiver apply complex
beamforming vectors $\vbf_{tx} \in \C^{n_{tx}}$ and $\vbf_{rx} \in \C^{n_{rx}}$ respectively.
We will assume these vectors are normalized to unity:
$\|\vbf_{tx}\|=\|\vbf_{rx}\|=1$.
Apply these beamforming vectors will reduce the MIMO channel $\Hbf$ in \eqref{eq:Hsum}
to an effective SISO channel with gain given by
\[
    G(\vbf_{tx},\vbf_{rx},\Hbf) = |\vbf_{rx}^*\Hbf\vbf_{tx}|^2.
\]
The maximum value for this gain would be
\[
    G_{\rm inst}(\Hbf) = \max_{\|\vbf_{tx}\|=\| \vbf_{rx}\|=1}
        G(\vbf_{tx},\vbf_{rx},\Hbf),
\]
and is found from the left and right singular vectors of $\Hbf$.
We can evaluate the average value of this gain as a ratio:
\beq \label{eq:bfgainInst}
    \mbox{BFGain}_{\rm inst} := 10\log_{10}\left[ \frac{\Exp G_{\rm inst}(\Hbf)}{
    G_{\rm omni}} \right],
\eeq
where we have compared the gain with beamforming to the omnidirectional gain
\beq \label{eq:Gomni}
    G_{\rm omni} := \frac{1}{n_{rx}n_{tx}} \Exp \|\Hbf\|^2_F,
\eeq
and the expectations in \eqref{eq:bfgainInst} and \eqref{eq:Gomni}
be taken over the small scale fading parameters in \eqref{eq:Hsum},
holding the large-scale fading parameters constant.
The ratio \eqref{eq:bfgainInst} represents the maximum increase in the gain
(effective decrease in path loss) from optimally steering the TX and RX
beamforming vectors.  It is easily verified that this gain is bounded by
\beq \label{eq:bfgainBnd}
    \mbox{BFGain}_{\rm inst} \leq 10\log_{10}(n_{rx}n_{tx}),
\eeq
with equality when $\Hbf$ in \eqref{eq:Hsum} is rank one
-- that is, there is no angular dispersion and the energy is concentrated in a
single direction.  In mmW systems, if the gain bound \eqref{eq:bfgainBnd}
can be achieved, the gain would be large:  for example,
if $n_{tx}=64$ and $n_{rx}=16$, the maximum gain in \eqref{eq:bfgainBnd}
is $10\log_{10}((64)(16)) \approx 30$ dB.
We call the gain in \eqref{eq:bfgainInst} the \emph{instantaneous gain}
since it represents the gain when the TX and RX beamforming vectors can
be selected based on the instantaneous small-scale fading realization of the channel,
and thus requires
CSI at both the TX and RX.  As described
above, such instantaneous beamforming may not be feasible.

We therefore consider an alternative and more conservative approach
known as \emph{long-term}
beamforming as described in \cite{Lozano:07}.  In long-term
beamforming, the TX and RX adapt the beamforming
vectors to the large-scale parameters (which are relatively slowly varying)
but not the small-scale ones.  One approach is to simply align the
TX and RX beamforming directions to the maximal eigenvectors of the covariance matrices,
\beq \label{eq:Qdef}
    \Qbf_{rx} := \Exp\left[ \Hbf\Hbf^* \right], \quad
    \Qbf_{tx} := \Exp\left[ \Hbf^*\Hbf \right],
\eeq
where the expectations are taken
with respect to the small-scale fading parameters assuming the large-scale
parameters are constant.
Since the small-scale fading is averaged out, these covariance matrices
are coherent over much longer periods of time and can be estimated
much more accurately.

When the beamforming vectors are held constant over the small-scale fading, we obtain
a SISO Rayleigh fading channel with an average gain of $\Exp G(\vbf_{tx},\vbf_{rx},\Hbf)$,
where the expectation is again taken over the small-scale fading.
We can define the \emph{long-term beamforming gain} as the ratio between the average
gain with beamforming and the average omnidirectional gain in \eqref{eq:Gomni},
\beq \label{eq:bfgainLong}
    \mbox{BFGain}_{\rm long} =
    10\log_{10}\left[ \frac{\Exp G(\vbf_{tx},\vbf_{rx},\Hbf)}{G_{\rm omni}} \right],
\eeq
where the beamforming vectors $\vbf_{tx}$ and $\vbf_{rx}$
are selected from the maximal eigenvectors of
the covariance matrices $\Qbf_{rx}$ and $\Qbf_{tx}$.

The long-term beamforming gain \eqref{eq:bfgainLong}
will be less than the instantaneous gain \eqref{eq:bfgainInst}.
To simplify the calculations,
we can approximately evaluate the long-term beamforming gain \eqref{eq:bfgainLong},
assuming a well-known
Kronecker model \cite{kermoal2002stochastic,mcnamara2002spatial},
\beq \label{eq:HKron}
    \Hbf \approx \frac{1}{\mathop{Tr}(\Qbf_{rx})}\Qbf_{rx}^{1/2} \Pbf \Qbf_{tx}^{1/2},
\eeq
where $\Pbf$ is an i.i.d.\ matrix with complex
Gaussian zero mean, unit variance components.
Under this approximate model,
it is easy to verify that the gain \eqref{eq:bfgainLong} is given by
the sum
\beq \label{eq:bfGainProd}
    \mbox{BFGain}_{\rm long} \approx \mbox{BFGain}_{TX} + \mbox{BFGain}_{RX},
\eeq
where the RX and TX beamforming gains are given by
\begin{subequations} \label{eq:bfGainRxTx}
\beqa \label{eq:bfGainRx}
    \mbox{BFGain}_{RX} &=& 10\log_{10} \left[ \frac{\lambda_{max}(\Qbf_{rx})}{(1/n_{rx})
        \sum_i \lambda_i(\Qbf_{rx})}
        \right] \\
    \mbox{BFGain}_{TX} &=& 10\log_{10} \left[ \frac{\lambda_{max}(\Qbf_{tx})}{(1/n_{tx})
        \sum_i \lambda_i(\Qbf_{tx})}
        \right],
\eeqa
\end{subequations}
where $\lambda_i(\Qbf)$ is the $i$-th eigenvalue of $\Qbf$ and
$\lambda_{max}(\Qbf)$ is the maximal eigenvalue.

\begin{figure}
    \centering
    \includegraphics[width=3.5in]{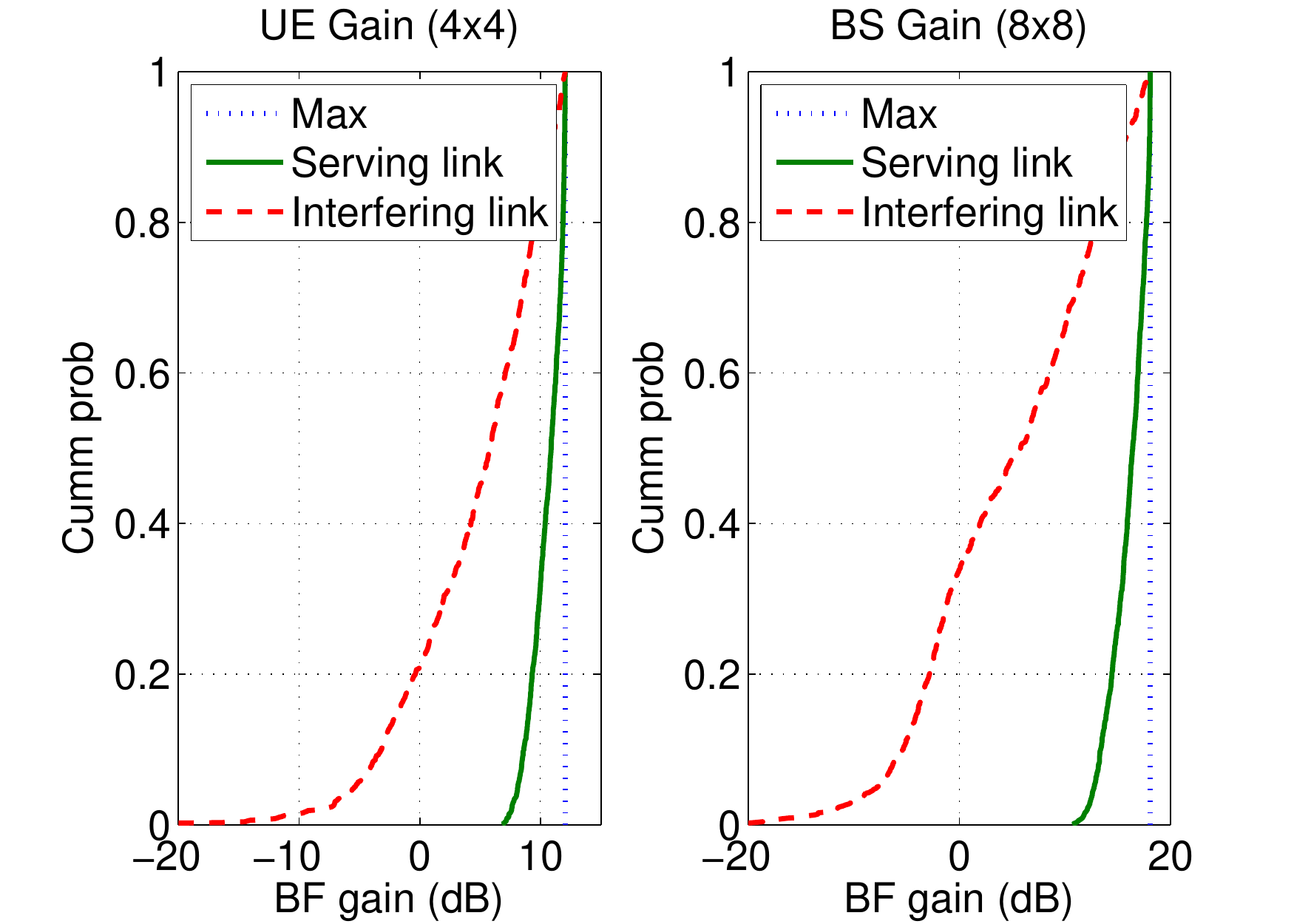}
    \caption{Distributions of the BS and UE long-term beamforming gains
    based on the 28~GHz models. The}   \label{fig:bfGain}
\end{figure}

Fig.~\ref{fig:bfGain}
plots the distributions of the long-term beamforming gains for the
UE and BS using the experimentally-derived channel model for 28~GHz
along with \eqref{eq:bfGainRxTx} (Note that BFGain$_{RX}$ and BFGain$_{TX}$
can be used for the either the BS or UE -- the gains are the same in either
direction).
In this figure, we have assumed a half-wavelength
8x8 uniform planar array at the BS transmitter and 4x4 uniform planar array
at the UE receiver.
The beamforming gains are random quantities since they depend on the large-scale
channel parameters.  The distribution of the beamforming gains
at the TX and RX along the serving links are shown in Fig.~\ref{fig:bfGain}
in the curves labeled ``Serving links".
Since we have assumed $n_{rx}=4^2 = 16$ antennas and $n_{tx}=8^2=64$ antennas,
the maximum beamforming gains possible would be 12 and 18 dB respectively,
and we see that long-term beamforming is typically able to get within 2-3 dB
of this maximum.  The average gain for instantaneous beamforming will be somewhere between the
long-term beamforming curve and the maximum value, so we conclude that loss from
long-term beamforming with respect to instantaneous beamforming is typically
bounded by 2-3 dB at most.

Also plotted in Fig.~\ref{fig:bfGain} is the distribution of the typical gain
along an interfering link.  This interfering gain provides a measure
of how directionally isolated a typical interferer will be.
The gain is estimated by selecting the beamforming direction from
a typical second-order matrix $\Qbf_{rx}$ or $\Qbf_{tx}$ and then
applying that beamforming direction onto a random second-order gain
with the same elevation angles.  The same elevation angles are used since the
BSs will likely have the same height.  We see that
the beamforming gains along these interfering directions is significantly
lower.  The median interfering beamforming gain
is approximately 6 dB lower in the RX and 9 dB in the TX.  This difference in gains
suggests that beamforming in mmW systems will be very effective in achieving a high
level of directional isolation.

Although the plots were shown for 28~GHz, very similar curves were observed at
73~GHz.

\subsection{Spatial Degrees of Freedom} \label{sec:spatDoF}

\begin{figure}
    \centering
    \includegraphics[width=3in]{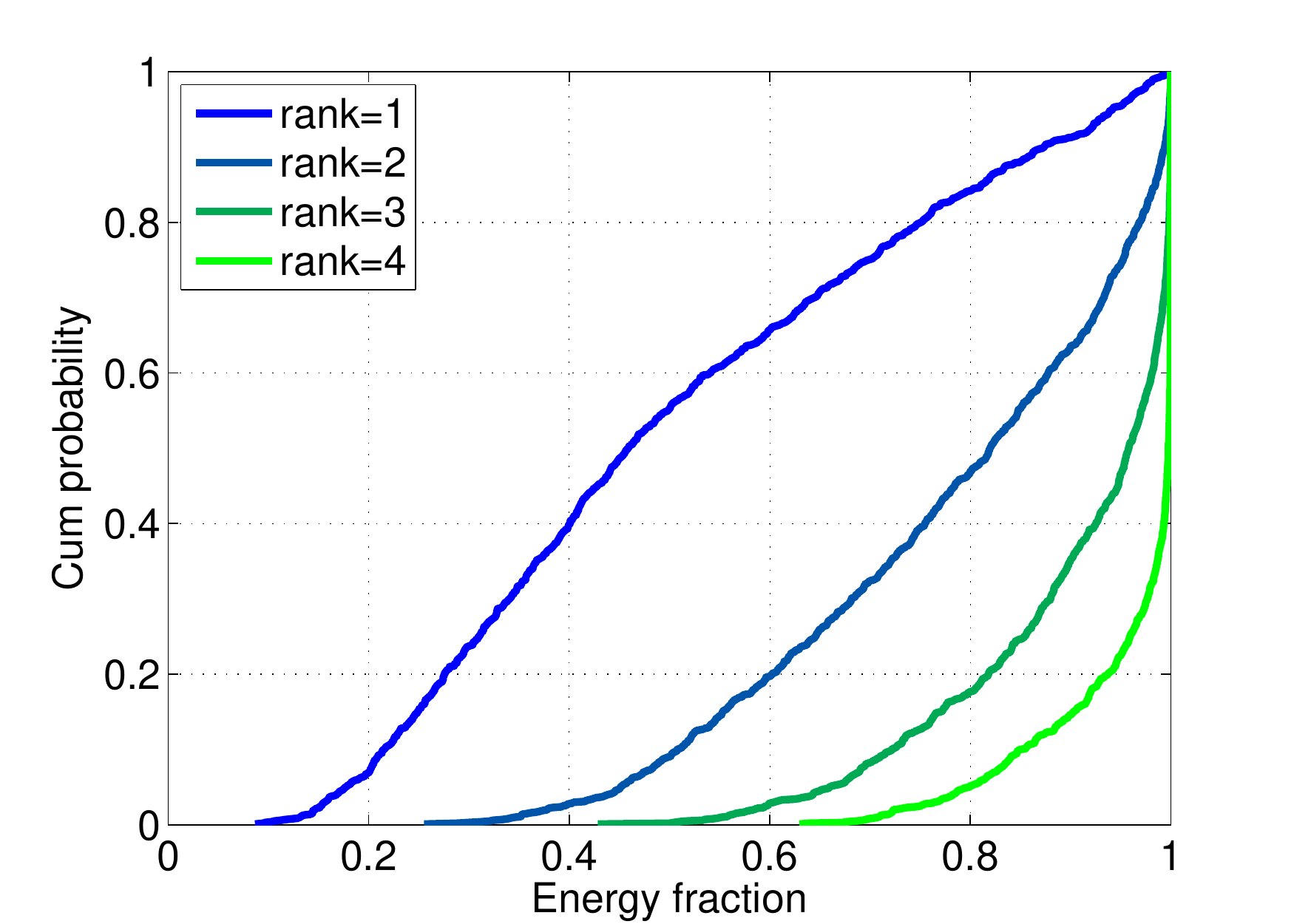}
    \caption{Distribution of the energy fraction
    in $r$ spatial directions for the 28~GHz channel model. }
    \label{fig:powFracRank}
\end{figure}

A second useful statistic to analyze is the typical rank of the channel.
The fact that we observed multiple path clusters between each TX-RX location pair
indicates the possibility of gains from spatial multiplexing \cite{TseV:07}.
To assess the amount of energy in multiple spatial streams, define
\[
    \phi(r) := \frac{1}{\Exp \|\Hbf\|^2_F}
        \max_{\Vbf_{rx}, \Vbf_{tx}}
        \Exp \| \Vbf_{rx}^*\Hbf\Vbf_{tx} \|_F^2,
\]
where the maximum is over matrices $\Vbf_{rx}  \in \C^{n_{rx}\x r}$ and
$\Vbf_{tx}  \in \C^{n_{tx}\x r}$
with $\Vbf_{rx}^*\Vbf_{rx}=I_r$ and $\Vbf_{tx}^*\Vbf_{tx}=I_r$.
The quantity $\phi(r)$ represents the fraction of energy that
can be captured by precoding onto an optimal $r$-dimensional
subspace at both the RX and TX.
Under the Kronecker model approximation \eqref{eq:HKron},
a simple calculation shows that this power fraction is given by the $r$ largest eigenvalues,
\[
    \phi(r) = \left[ \frac{\sum_{i=1}^r \lambda_i(\Qbf_{rx})}{
        \sum_{i=1}^{n_{rx}}\lambda_i(\Qbf_{rx})}\right]
        \left[ \frac{\sum_{i=1}^r \lambda_i(\Qbf_{tx})}{
        \sum_{i=1}^{n_{tx}}\lambda_i(\Qbf_{tx})}\right],
\]
where $\Qbf_{rx}$ and $\Qbf_{tx}$ iare the spatial covariance matrices \eqref{eq:Qdef} and
$\lambda_i(\Qbf)$ is the $i$-th largest eigenvalue of $\Qbf$.
Since the power fraction is dependent on the second-order, long-term
channel statistics, it is a random variable.  Fig.~\ref{fig:powFracRank} plots
the distribution of $\phi(r)$ for values $r=1,\ldots,4$ for the
experimentally-derived 28 GHz channel model.  The power fractions for the 73~GHz
are not plotted, but are similar.

If the channel had  no angular dispersion per cluster,
then $\Qbf_{rx}$ and $\Qbf_{tx}$ would have rank one and all the energy could be
captured with one spatial dimension, i.e.\ $\phi(r)=1$ with $r=1$.
However, since the channels have possibly multiple clusters and the
clusters have a non-zero angular dispersion, we see that there is significant
energy in higher spatial dimensions.  For example,
Fig.~\ref{fig:powFracRank} shows that in the median channel,
a single spatial dimension is only able to capture approximately 50\% of the channel energy.
Two degrees of freedom are needed to capture the 80\% of the channel
energy and three dimensions are needed for 95\%.  These numbers suggest that many
locations will be capable of providing single-user MIMO gains with two and even
three streams.  Note that further spatial degrees of freedom are possible
with multi-user MIMO beyond the rank of the channel to any one user.

\section{Capacity Evaluation}
\subsection{System Model}
\begin{table}
\caption{Default network parameters}
\label{table:para}
\hfill{}
 \begin{tabular}{|>{\raggedright}p{1.25in}|>{\raggedright}p{1.75in}|}
	\hline
	\textbf{Parameter}  &  \textbf{Description} \tabularnewline \hline
	BS layout and sectorization &
        Hexagonally arranged cell sites placed in a 2km x 2km square area
        with three cells per site. \tabularnewline \hline
	UE layout   &   Uniformly dropped in area with average of 10 UEs per
        BS cell (i.e.\ 30 UEs per cell site).
        \tabularnewline \hline
	Inter-site distance (ISD) &    200 m \tabularnewline \hline
    Carrier frequency & 28 and 73 GHz \tabularnewline \hline
    Duplex mode & TDD \tabularnewline \hline
	Transmit power  &  20 dBm (uplink), 30 dBm (downlink) \tabularnewline \hline
	Noise figure  &  5~dB (BS), 7~dB (UE) \tabularnewline \hline
    BS antenna & 8x8 $\lambda/2$ uniform planar array \tabularnewline \hline
    UE antenna & 4x4 $\lambda/2$ uniform planar array for 28~GHz and
          8x8 array for 73~GHz.  \tabularnewline \hline
    Beamforming & Long-term, single stream \tabularnewline \hline
  \end{tabular}
\hfill{}
\end{table}

To assess the system capacity under the experimentally-measured
channel models, we follow a standard cellular evaluation methodology \cite{3GPP36.814}
where the BSs and UEs
are randomly ``dropped" according to some statistical model and the performance
metrics are then measured over a number of random realizations of the network.
Since we are interested in small cell networks, we follow a BS and UE distribution
similar to the 3GPP Urban Micro (UMi) model in \cite{3GPP36.814} with some
parameters taken from the Samsung mmW study~\cite{KhanPi:11,KhanPi:11-CommMag}.
The specific parameters are shown in Table~\ref{table:para}.
Similar to 3GPP UMi model, the BS cell sites are distributed
 in a uniform hexagonal pattern with three cells (sectors) per site
 covering a 2~km by
 2 km area with an inter-site distance (ISD) of 200 m.  This layout
 leads to 130 cell sites
 (390 cells) per drop.
UEs are uniformly distributed over the area at a density of 10 UEs per cell -- which
also matches the 3GPP UMi assumptions.
The maximum transmit power of 20~dBm at the UE and
30~dBm are taken from~\cite{KhanPi:11,KhanPi:11-CommMag}.
Note that since our channel models were based on data from
receivers in outdoor locations,
implicit in our model is that all users are outdoors.
If we included mobiles that were indoor, it is likely that the capacity numbers
would be significantly lower since mmW signals cannot penetrate many building
materials.

These transmit powers
are reasonable since current CMOS RF power amplifiers in the mmW range
exhibit peak efficiencies of at least 8\% \cite{Z3,Z4}.
This implies that the UE TX power
of 20~dBm and BS TX power of 30 dBm can be achieved with powers of 1.25W and 12.5W,
respectively.

\subsection{Beamforming Modeling}

Although our preliminary calculations in Section~\ref{sec:spatDoF} suggest
that the channel may support spatial multiplexing,
we consider only single stream processing where the RX and TX
beamforming is designed to maximize SNR without regard to interference.
That is, there is no interference nulling.
It is possible that  more
 advanced techniques such as inter-cell coordinated beamforming and MIMO spatial multiplexing \cite{ZhangMadhow1,Heath:partialBF} may offer further
 gains, particularly for mobiles close to the cell.  Indeed, as we saw in Section
 ~\ref{sec:spatDoF}, many UEs have at least two significant spatial degrees of freedom
 to support single user MIMO.  Multiuser MIMO and SDMA may offer even greater
 opportunities for spatial multiplexing.  However, modeling of MIMO and SDMA, particularly
 under constraints on the number of spatial streams requires further work and will be studied in upcoming
 papers.

Under the assumption of signal stream processing,
the link between each TX-RX pair can be modeled as an effective
single-input single-output (SISO) channel with an effective path loss
that accounts for the total power received on the different path clusters
between the TX and RX
and the beamforming applied at both ends of the link.
The beamforming gain will be distributed following the distributions in
Section~\ref{sec:bfGain}.

\subsection{MAC Layer Assumptions}

Once the effective path losses are determined between all TX-RX pairs,
we can compute the average SINR at each RX.  The SINR in turn determines
the rate per unit time and bandwidth allocated to the mobile.
In an actual cellular system, the achieved rate (goodput) will depend on
the average SNR through a number of factors including
the channel code performance, channel quality indicator (CQI) reporting, rate adaptation
and Hybrid automatic repeat request (HARQ) protocol. In this work, we abstract this process and assume a simplified, but widely-used,
model \cite{MogEtAl:07}, where the spectral efficiency is assumed to be given by the Shannon
capacity with some loss $\Delta$:
\beq \label{eq:rateSNR}
    \rho = \min\left\{ \log_2\left(1 + 10^{0.1(\SNR-\Delta)}\right), \rho_{max}\right\},
\eeq
where $\rho$ is the spectral efficiency in bps/Hz,
the SNR and loss factor $\Delta$ are in dB,
and $\rho_{max}$ is the maximum
spectral efficiency.  Based on analysis of current LTE turbo codes,
the paper \cite{MogEtAl:07} suggests parameters
$\Delta = 1.6$ dB and $\rho_{max} = 4.8$ bps/Hz.
Assuming similar codes can be used for a mmW system,
we apply the same $\rho_{max}$ in this simulation, but increase $\Delta$ to 3 dB to account for fading.  This increase in $\Delta$ is necessary since the results
in \cite{MogEtAl:07} are based on AWGN channels.  The 1.4~dB increase used here
is consistent with results from link error prediction methods such as
\cite{ikuno2011novel}.
Note that all rates stated in this paper \emph{do not} include the half duplex loss, which must be added depending on the UL-DL ratio.
The one exception to this accounting is the comparison in
Section~\ref{sec:capResult}
between mmW and LTE systems, where we explicitly assume a 50-50 UL-DL duty cycle.

For the uplink and downlink scheduling,
we use proportional fair scheduling with full buffer traffic.
Since we assume that we cannot exploit multi-user diversity and only
schedule on the average channel conditions, the proportional fair
assumption implies that each UE will get an equal fraction of the time-frequency
resources.
In the uplink, we will additionally assume that the multiple access scheme
enables multiple UEs to be scheduled at the same time.  In OFDMA systems
such as LTE, this can be enabled by scheduled the UEs on different
resource blocks.  Enabling multiple UEs to transmit at the same time
provides a significant power boost.
However, supporting such multiple access also requires that the BS
can receive multiple simultaneous beams.  As mentioned above, such reception would require
multiple RF chains at the BS, which will add some complexity and power consumption.
Note, however, that all processing in this study, requires only single streams at the mobile, which
is the node that is more constrained in terms of processing power.

\subsection{Uplink and Downlink Throughput} \label{sec:capResult}

\begin{figure}
    \centering
    \includegraphics[width=3in]{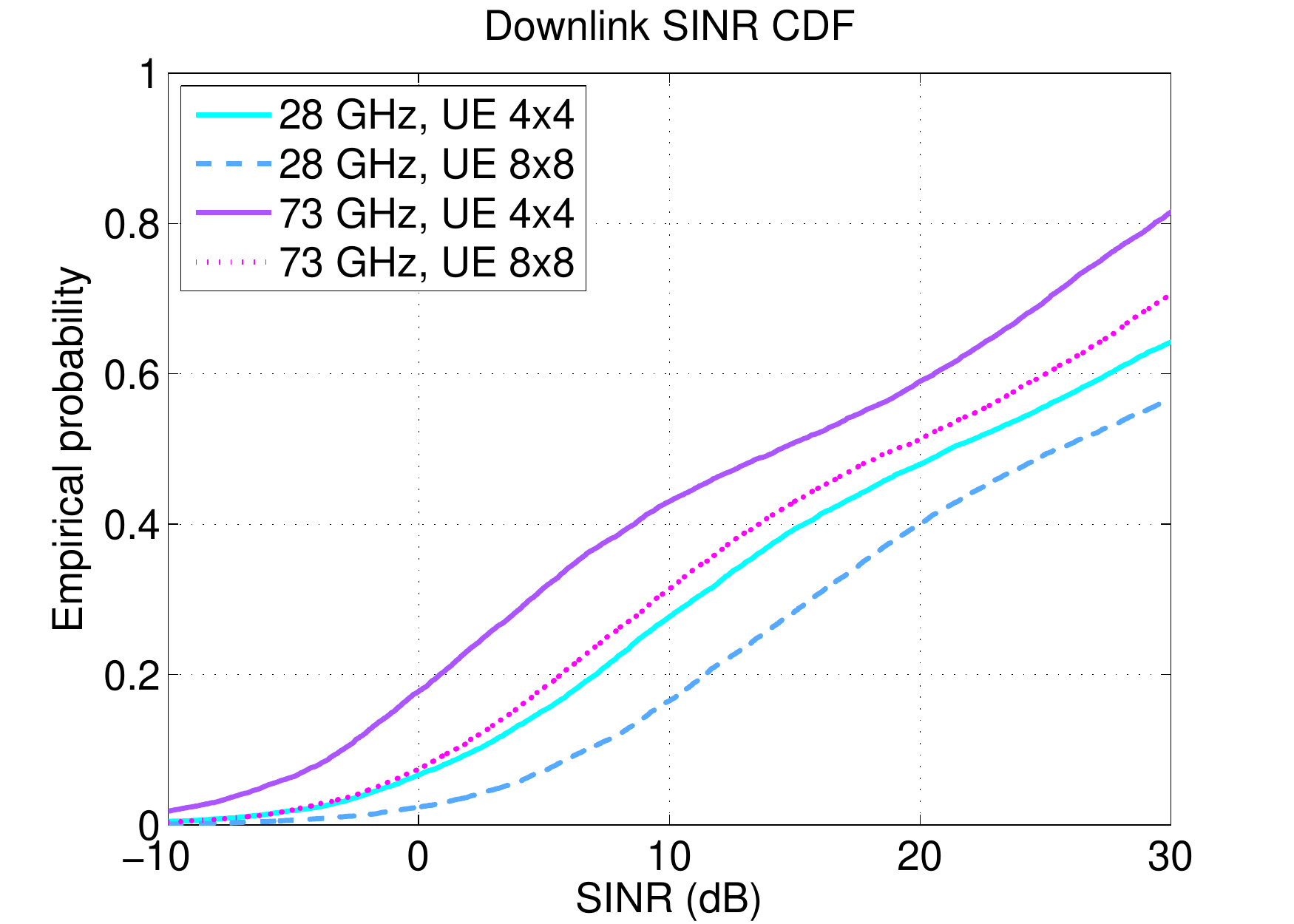}
    \includegraphics[width=3in]{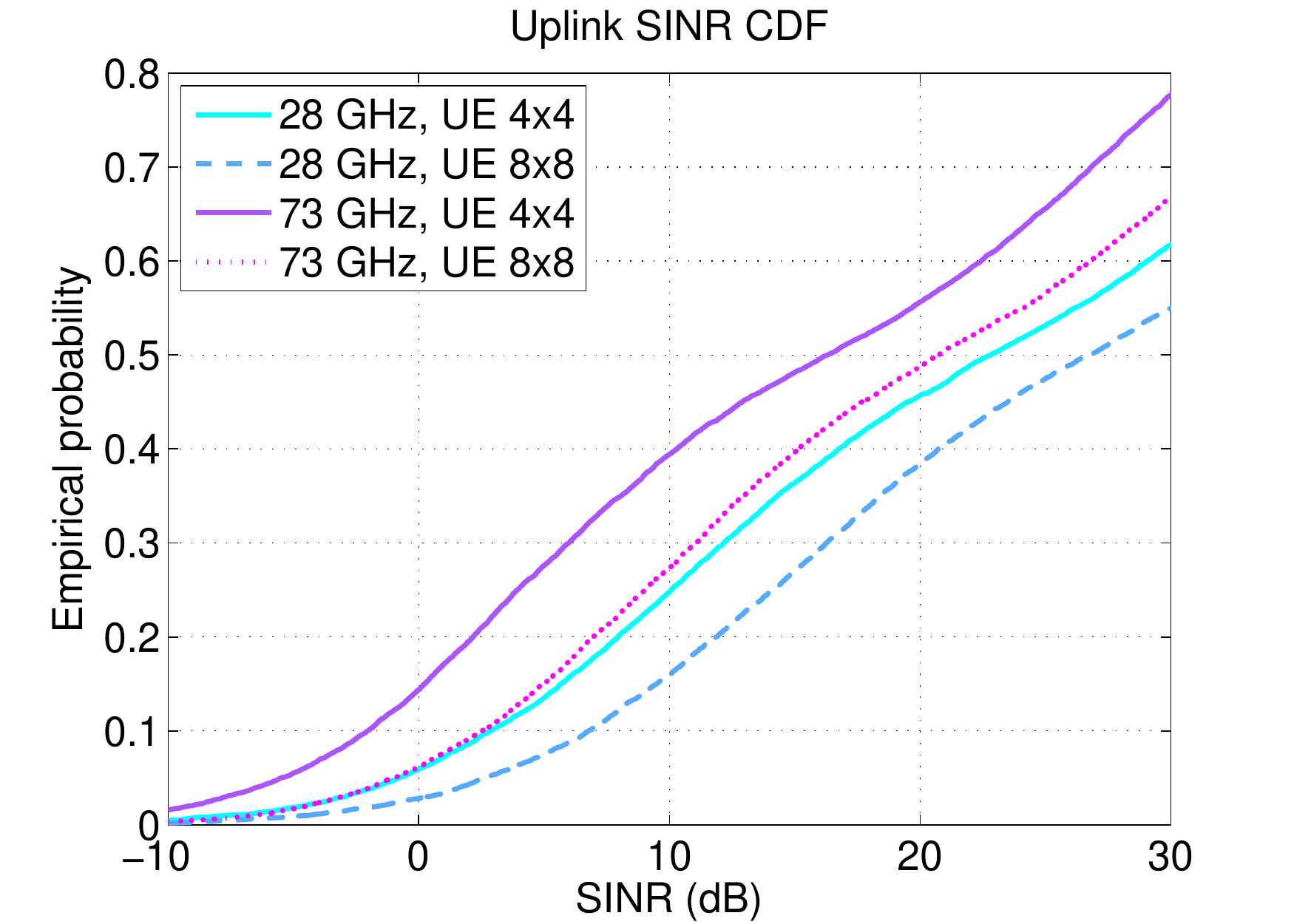}
    \caption{Downlink (top plot) / uplink (bottom plot) SINR CDF
        at 28 and 73~GHz with 4x4 and 8x8 antenna arrays at the UE.
        The BS antenna array is held at 8x8.}
    \label{fig:sinrGeoTxPow}
\end{figure}

\begin{figure}
    \centering
    \includegraphics[width=3in]{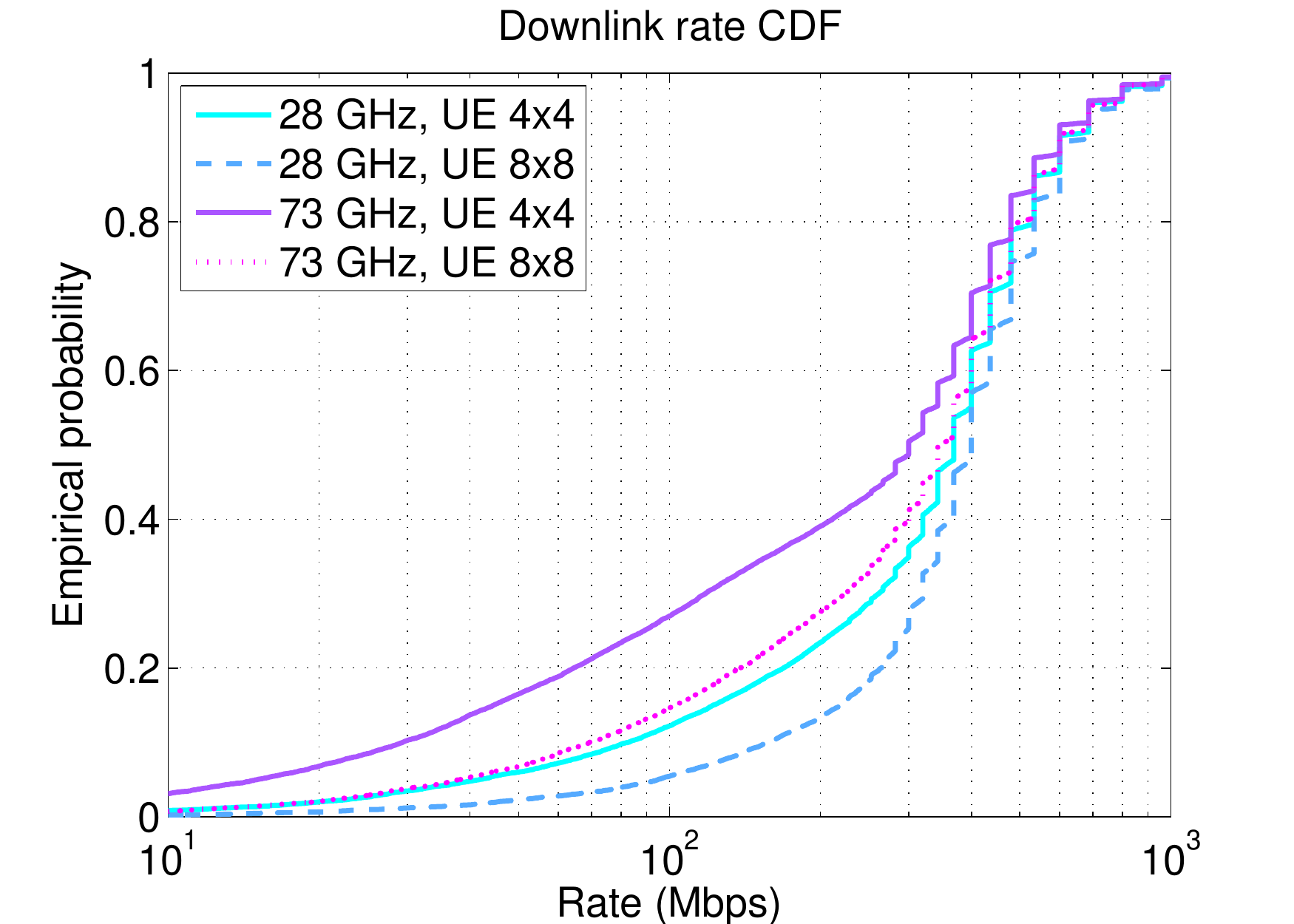}
    \includegraphics[width=3in]{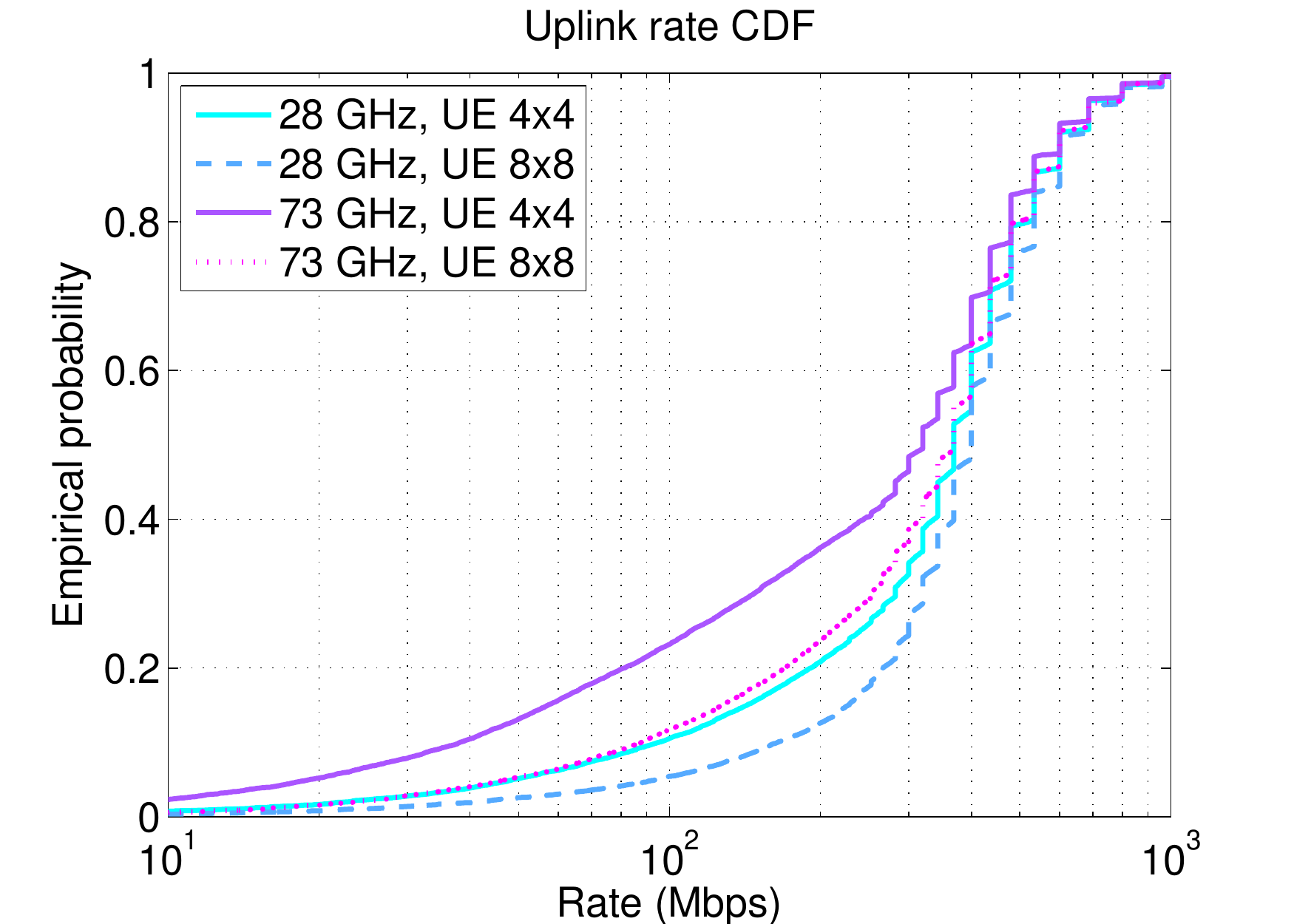}
    \caption{Downlink (top plot) / uplink (bottom plot) rate CDF
    at 28 and 73~GHz with 4x4 and 8x8 antenna arrays at the UE.
    The BS antenna array is held at 8x8.}
    \label{fig:rateGeoTxPow}
\end{figure}

We plot SINR and rate distributions in Figs.~\ref{fig:sinrGeoTxPow} and
\ref{fig:rateGeoTxPow} respectively. The distributions are plotted for
both 28 and 73~GHz and for 4x4 and 8x8 arrays at the UE.  The BS antenna
array is held at 8x8 for all cases.    There are a few important observations
we can make.

First, for the same number of antenna elements, the rates for 73~GHz
are approximately half the rates for the 28~GHz.  However, a 4x4 $\lambda/2$-array
at 28~GHz would take about the same area as an 8x8 $\lambda/2$ array at 73~GHz.
Both would be roughly 1.5$\times$ 1.5 cm$^2$,
which could be easily accommodated in a handheld
mobile device.  In addition, we see that 73~GHz 8x8 rate and SNR distributions
are very close to the 28~GHz 4x4 distributions, which is reasonable since we are
keeping the antenna size constant.  Thus, we can conclude that the
loss from going to the higher frequencies can be made up from larger numbers of antenna
elements without increasing the physical antenna area.

As a second point, we can compare the SINR distributions in Fig.~\ref{fig:sinrGeoTxPow}
to those of a traditional cellular network.
Although the SINR distribution for a cellular network in a traditional frequency
is not plotted here, the SINR distributions in Fig.~\ref{fig:sinrGeoTxPow}
are actually slightly better than those found in cellular
evaluation studies \cite{3GPP36.814}.
For example,  in Fig.~\ref{fig:sinrGeoTxPow}, only about 5 to 10\% of the mobiles
appear under 0 dB, which is a lower fraction than typical cellular deployments.
We conclude that, although mmW systems have an omnidirectional path loss that
is 20 to 25 dB worse than conventional microwave frequencies, short cell radii combined
with highly directional beams are able to completely compensate for the loss.


As one final point, Table \ref{table:se} provides a comparison of mmW and current LTE systems.
The LTE capacity numbers are taken from the average of industry
reported evaluations given in~\cite{3GPP36.814}
-- specifically Table 10.1.1.1-1 for the downlink
and Table 1.1.1.3-1 for the uplink.  The LTE evaluations include
advanced techniques such as SDMA, although not coordinated multipoint.
For the mmW capacity, we assumed 50-50 UL-DL TDD split and a 20\%
control overhead in both the UL and DL directions.
Note that in the spectral efficiency numbers for the mmW system,
we have included the 20\% overhead, but not the 50\% UL-DL split.
Hence the cell throughput is given by $C=0.5\rho W$, where $\rho$ is the
spectral efficiency, $W$ is the bandwidth, and the 0.5 accounts for the
duplexing.

Under these assumptions, we see that the mmW system for either the 28~GHz
4x4 array or 73~GHz 8x8 array provides a significant $> 25$-fold
increase of overall cell throughput over the LTE system.
Of course, most of the gains are simply coming from the increased spectrum:
the operating bandwidth of mmW is chosen as 1~GHz as opposed to 20+20 MHz in LTE
-- so the mmW system has 25 times more bandwidth.
However, this is a basic mmW system with no spatial multiplexing or other advanced
techniques --  we expect even higher gains when advanced technologies are applied to optimize the mmW system.
While the lowest 5\% cell edge rates are less dramatic, they still offer a 10 to 13
fold increase over the LTE cell edge rates.

\begin{table*}
\caption{mmW and LTE cell throughput/cell edge rate comparison.}
\label{table:se}
\hfill{}
\begin{threeparttable}
 \begin{tabular}{
     |>{\raggedright}p{0.8in}|>{\raggedright}p{0.9in}|
     	>{\raggedright}p{0.3in}|>{\raggedright}p{1.0in}|
      >{\raggedright}p{0.2in}|>{\raggedright}p{0.2in}|
      >{\raggedright}p{0.4in}|>{\raggedright}p{0.4in}|
      >{\raggedright}p{0.3in}|>{\raggedright}p{0.3in}|}
	\hline

	\multirow{2}{0.5in}{System}& System Bandwidth & UE ant & \multirow{2}{1.1in}{NLOS-LOS-Outage model}
     &   \multicolumn{2}{c|}{\multirow{2}{0.4in}{Spec. eff (bps/Hz)}} &
      \multicolumn{2}{c|}{\multirow{2}{0.8in}{Cell throughput (Mbps/cell)}}
    &  \multicolumn{2}{c|}{\multirow{2}{0.8in}{5\%~Cell~edge~rate (Mbps/UE)}}
        \tabularnewline \cline{5-10}
   & & & & DL & UL & DL & UL & DL & UL \tabularnewline \hline
	\multirow{5}{0.8in}{28~GHz~mmW} & \multirow{5}{0.8in}{1~GHz TDD}
	   & 8x8 &  Hybrid	& 3.34 & 3.16 & 1668 & 1580 & 52.28 & 34.78
 \tabularnewline \cline{3-10}
     & & \multirow{4}{0.5in}{4x4} & Hybrid & 3.03 & 2.94 & 1514 & 1468 & 28.47 & 19.90
     \tabularnewline \cline{4-10}
     & & & Hybrid, $d_\mathrm{shift}$~=~50m &  2.90 & 2.91 & 1450 & 1454 & 17.62 & 17.49
     \tabularnewline \cline{4-10}
     & & & Hybrid, $d_\mathrm{shift}$~=~75m &  2.58 & 2.60 & 1289 & 1298 & 0.54 & 0.09
     \tabularnewline \cline{4-10}
     & & & No LOS, $d_\mathrm{shift}$~=~50m & 2.16 & 2.34 & 1081 & 1168 & 11.14 & 15.19
        \tabularnewline \hline
  \multirow{2}{0.8in}{73~GHz~mmW} & \multirow{2}{0.8in}{1~GHz TDD}
	   & 4x4 & Hybrid & 2.58 & 2.58 & 1288 & 1291 & 10.02 & 8.92
     \tabularnewline \cline{3-10}
 & & 8x8 & Hybrid & 2.93 & 2.88 & 1465 & 1439 & 24.08 & 19.76
 		 \tabularnewline \hline
    2.5~GHz~LTE & 20+20 MHz FDD
     & 2 & & 2.69 & 2.36 & 53.8 & 47.2 & 1.80 & 1.94
     \tabularnewline \hline
  \end{tabular}
  \begin{tablenotes}
   \item Note 1. Assumes 20\% overhead, 50\% UL-DL duty cycle and 8x8 BS antennas for the mmW system
   \item Note 2. Assumes 2 TX 4 RX antennas at BS side for LTE system
   \item Note 3. Long-term, non-coherent beamforming are assumed at both the
   BS and UE in the mmW system.
   However, the mmW results assume no spatial multiplexing gains, whereas the LTE results from
   \cite{3GPP36.814} include spatial multiplexing and beamforming.
    \end{tablenotes}
    \end{threeparttable}
\hfill{}
\end{table*}

\subsection{Directional Isolation}

In addition to the links being in a relatively high SINR,
an interesting feature of mmW systems is that thermal noise dominates
interference.
Although the distribution of the interference to noise ratio is not plotted,
we observed that in 90\% of the links, thermal noise was larger than the interference
-- often dramatically so.  We conclude that highly directional transmissions
used in mmW systems combined with short cell radii
result in links that are in relatively high SINR with
little interference.  This feature is in stark contrast to current dense cellular
deployments where links are overwhelmingly interference-dominated.

\subsection{Effect of Outage}

\begin{figure}
    \centering
    \includegraphics[width=3in]{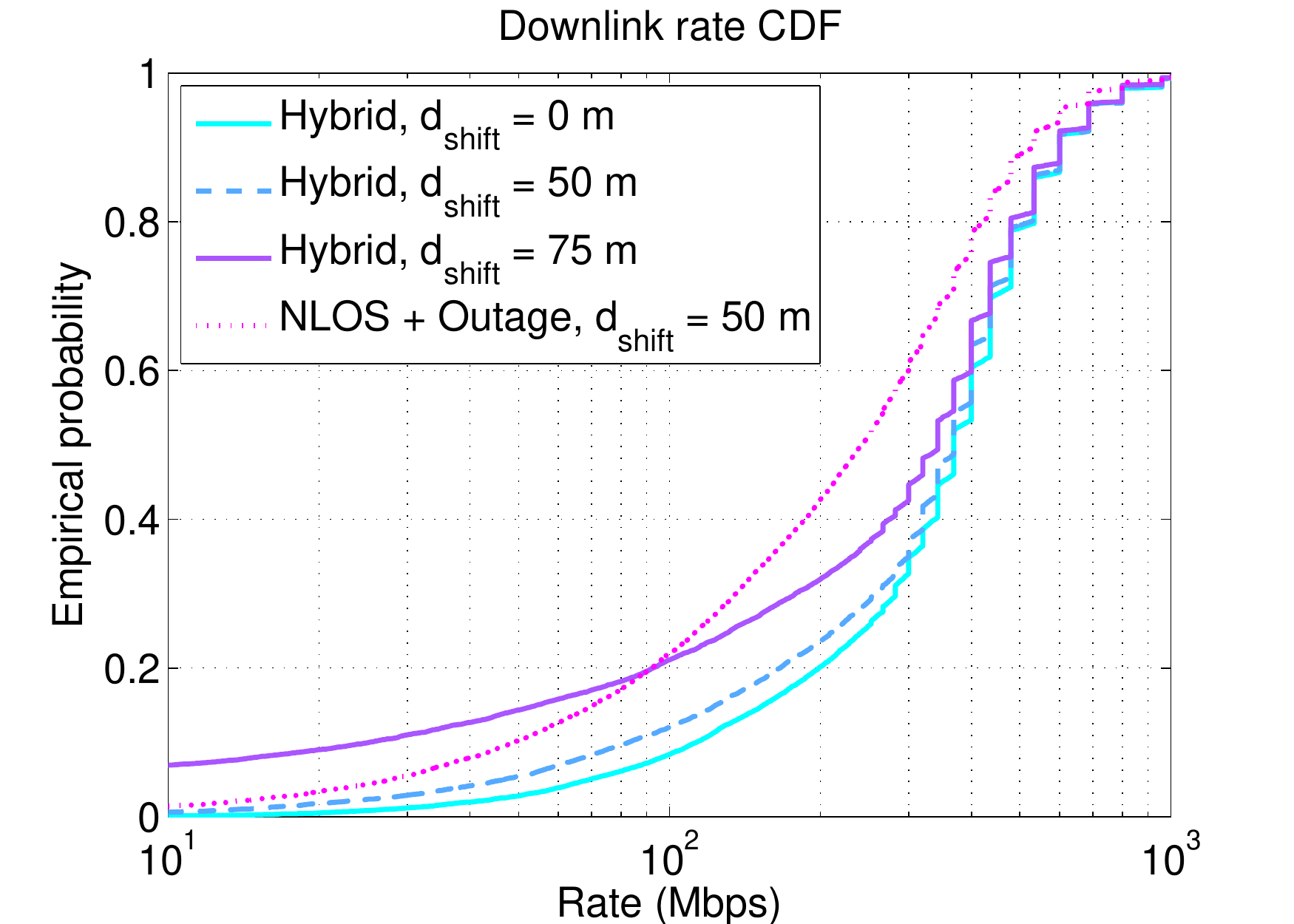}
    \includegraphics[width=3in]{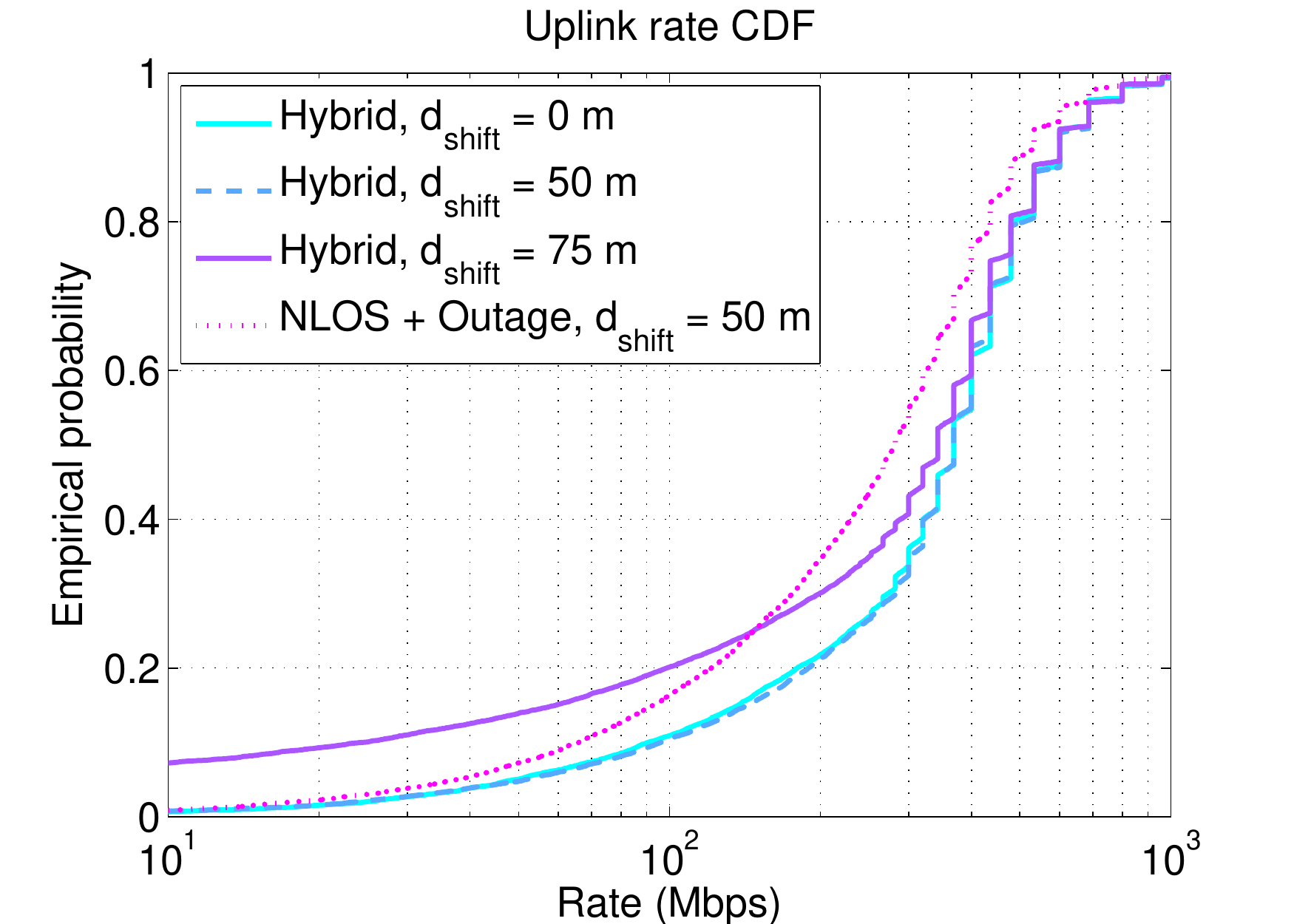}
    \caption{Downlink (top plot) / uplink (bottom plot) rate CDF
        under the link state model with various parameters.  The carrier frequency is 28~GHz. $d_\mathrm{shift}$ is the amount by which the outage curve in \eqref{eq:Pout} is shifted to the left.
    }
    \label{fig:rateBlk}
\end{figure}

One of the significant features of mmW systems
is the presence of outage -- the fact that there is a non-zero probability
that the signal from a given BS can be completely blocked and hence
not detectable.
The parameters in the hybrid LOS-NLOS-outage model \eqref{eq:phybrid}
were based on our data in one region of NYC.  To understand the potential
effects of different outage conditions, Fig.~\ref{fig:rateBlk} shows
the distribution of rates under various NLOS-LOS-outage probability models.
The curve labeled ``hybrid, $d_{\rm shift}=0$" is the baseline model
with parameters provided on Table~\ref{tbl:largeScaleParam} that we have used up to now.
These are the parameters based on the fitting the NYC data.
This model is compared to two models with heavier outage
created by shifting $p_\out(d)$ to the left by 50~m and 75~m,
shown in the second and third curves.
The fourth curve labeled ``NLOS+outage, $d_{\rm shift}=50$ m"
uses the shifted outage and also removes all the LOS links -- hence all the links
are either in an outage or NLOS state.
In all cases, the carrier frequency is 28 GHz and the
we assumed a 4x4 antenna array at the UE.  Similar findings were observed
at 73 GHz and 8x8 arrays.

We see that, even with a 50~m shift in the outage curve (i.e.\ making
the outages occur 50~m closer than predicted by our model), the system performace
is not significantly affected. In fact, there is
a slight improvement in UL rates due to suppressed interference and only
slight decrease in DL cell throughput and edge rates  -- a point also observed in
\cite{bai2013analysis}.
However, when we increase the outage even more $d_{shift}=75$~m,
we start to see that many
UEs cannot establish a connection to any BS since the outage radius becomes comparable to the cell radius, which is 100~m. In other words, there is a non-zero
probability
that mobiles physically close to a cell may be in outage to that cell.
These mobiles will need
to connect to a much more distant cell. Therefore, we see the dramatic decrease in edge cell rate. Note that in our model,
the front-to-back antenna gains are assumed to infinite, so mobiles that are
blocked to one sector of a cell site cannot see any other sectors.

Fig.~\ref{fig:rateBlk} also shows that the throughputs are greatly benefitted
by the presence of LOS links.  Removing the LOS links so that all
links are in either an NLOS or outage states results in a significant drop in
rate.  However, even in this case, the mmW system offers a greater than 20 fold
increase in rate over the comparison LTE system.
It should be noted that the capacity numbers reported in \cite{RanRapE:14},
which were based on an earlier version of this paper, did not include any LOS links.

We conclude that, in environments with outages condition similar to, or even somewhat
worse than the NYC environment where our experiments were conducted,
the system will be very robust to outages.   This is extremely encouraging
since signal outage is one of the key concerns for the feasibility of mmW cellular
in urban environments.  However, should outages be dramatically worse than
the scenarios in our experiments (for example, if the outage radius is
shifted by 75~m), many mobiles will indeed lose connectivity even when they
are near a cell.
In these circumstances, other techniques such as relaying, more dense
cell placement or fallback to conventional frequencies will likely be needed.
Such ``near cell" outage will likely be present when mobiles are placed
indoors, or when
humans holding the mobile device
block the paths to the cells.  These factors were not considered in our
measurements, where receivers were placed at outdoor locations with no obstructions
near the cart containing the measurement equipment.

\section*{Conclusions}

We have provided the first detailed statistical mmW
channel models for several of the key channel parameters including
the path loss, and spatial characteristics and outage probability.
The models are based on real experimental data collected in New York City
in 28 and 73~GHz.
The models reveal that
signals at these frequencies
can be detected at least 100~m to 200~m from the potential cell sites,
even in absence of LOS connectivity. In fact, through building reflections,
signals at many locations
arrived with multiple path clusters to support spatial multiplexing
and diversity.

Simple statistical models, similar to those in current cellular standards such
as \cite{3GPP36.814} provide a good fit to the observations.
Cellular capacity evaluations based on these models predict an order of magnitude
increase in capacity over current state-of-the-art 4G systems under reasonable
assumptions on the antennas, bandwidth and beamforming.
These findings provide strong evidence for the viability of small cell
outdoor mmW systems even in challenging urban canyon
environments such as New York City.

The most significant caveat in our analysis is the fact that the measurements,
and the models derived from those measurements, are based on outdoor street-level
locations.  Typical urban cellular evaluations, however, place a large fraction
of mobiles indoors, where mmW signals will likely not penetrate.
Complete system evaluation with indoor mobiles will need further study.
Also, indoor locations and other coverage holes
may be served either via multihop relaying or fallback to conventional microwave
cells and further study will be needed to quantify the performance of these systems.

\section*{Acknowledgements}
The authors would like to deeply thank several students and colleagues
for providing the path loss data
\cite{Rappaport:12-28G,Rappaport:28NYCPenetrationLoss,Samimi:AoAD,rappaportmillimeter}
that made this research possible:
Yaniv Azar, Felix Gutierrez, DuckDong Hwang,
Rimma Mayzus, George McCartney, Shuai Nie,
Jocelyn K. Schulz,  Kevin Wang,
George N. Wong and  Hang Zhao.
This work also benefitted significantly from discussions with our
industrial partners in NYU WIRELESS program including Samsung, NSN, Qualcomm
and InterDigital.

\bibliographystyle{IEEEtran}
\bibliography{bibl}

\newcommand{\SortNoop}[1]{}
\begin{thebibliography}{10}
\providecommand{\url}[1]{#1}
\csname url@samestyle\endcsname
\providecommand{\newblock}{\relax}
\providecommand{\bibinfo}[2]{#2}
\providecommand{\BIBentrySTDinterwordspacing}{\spaceskip=0pt\relax}
\providecommand{\BIBentryALTinterwordstretchfactor}{4}
\providecommand{\BIBentryALTinterwordspacing}{\spaceskip=\fontdimen2\font plus
\BIBentryALTinterwordstretchfactor\fontdimen3\font minus
  \fontdimen4\font\relax}
\providecommand{\BIBforeignlanguage}[2]{{%
\expandafter\ifx\csname l@#1\endcsname\relax
\typeout{** WARNING: IEEEtran.bst: No hyphenation pattern has been}%
\typeout{** loaded for the language `#1'. Using the pattern for}%
\typeout{** the default language instead.}%
\else
\language=\csname l@#1\endcsname
\fi
#2}}
\providecommand{\BIBdecl}{\relax}
\BIBdecl

\bibitem{CiscoVNI:latest}
Cisco, ``{Cisco Visual Network Index}: Global mobile traffic forecast update,''
  2013.

\bibitem{EricssonMDT:latest}
Ericsson, ``Traffic and market data report,'' 2011.

\bibitem{UMTSForecast}
{UMTS Forum}, ``Mobile traffic forecasts: 2010-2020 report,'' in \emph{UMTS
  Forum Report}, vol.~44, 2011.

\bibitem{KhanPi:11}
F.~Khan and Z.~Pi, ``{Millimeter-wave {M}obile {B}roadband ({MMB}):
  {U}nleashing 3-300GHz Spectrum},'' in \emph{Proc.\ IEEE Sarnoff Symposium},
  Mar. 2011.

\bibitem{KhanPi:11-CommMag}
------, ``{An introduction to millimeter-wave mobile broadband systems},''
  \emph{IEEE Comm. Mag.}, vol.~49, no.~6, pp. 101 -- 107, Jun. 2011.

\bibitem{Ted:60Gstate11}
T.~S. Rappaport, J.~N. Murdock, and F.~Gutierrez, ``{State of the art in 60-GHz
  integrated circuits and systems for wireless communications},''
  \emph{Proceedings of the IEEE}, vol.~99, no.~8, pp. 1390 -- 1436, August
  2011.

\bibitem{PietBRPC:12}
P.~Pietraski, D.~Britz, A.~Roy, R.~Pragada, and G.~Charlton, ``Millimeter wave
  and terahertz communications: Feasibility and challenges,'' \emph{ZTE
  Communications}, vol.~10, no.~4, pp. 3--12, Dec. 2012.

\bibitem{BocHLMP:14}
F.~Boccardi, R.~W. {Heath, Jr.}, A.~Lozano, T.~L. Marzetta, and P.~Popovski,
  ``Five disruptive technology directions for {5G},'' to appear in \emph{IEEE
  Comm. Magazine}, 2014.

\bibitem{RanRapE:14}
S.~Rangan, T.~S. Rappaport, and E.~Erkip, ``Millimeter-wave cellular wireless
  networks: Potentials and challenges,'' \emph{Proceedings of the IEEE}, vol.
  102, no.~3, pp. 366--385, March 2014.

\bibitem{Doan:04}
C.~Doan, S.~Emami, D.~Sobel, A.~Niknejad, and R.~Brodersen, ``{Design
  considerations for 60 GHz CMOS radios},'' \emph{IEEE Comm. Mag.}, vol.~42,
  no.~12, pp. 132 -- 140, 2004.

\bibitem{Doan:05}
C.~Doan, S.~Emami, A.~Niknejad, and R.~Brodersen, ``Millimeter-wave {CMOS}
  design,'' \emph{IEEE J. Solid-State Circuts}, vol.~40, no.~1, pp. 144--155,
  2005.

\bibitem{ZhaLiu:09}
Y.-P. Zhang and D.~Liu, ``Antenna-on-{C}hip and {A}ntenna-in-{P}ackage
  solutions to highly integrated millimeter-wave devices for wireless
  communications,'' \emph{IEEE Trans.\ Antennas and Propagation}, vol.~57,
  no.~10, pp. 2830--2841, 2009.

\bibitem{gutierrez2009chip}
F.~Gutierrez, S.~Agarwal, K.~Parrish, and T.~S. Rappaport, ``On-chip integrated
  antenna structures in {CMOS} for 60 {GHz WPAN} systems,'' \emph{IEEE J. Sel.
  Areas Comm.}, vol.~27, no.~8, pp. 1367--1378, 2009.

\bibitem{Nsenga:10}
J.~Nsenga, A.~Bourdoux, and F.~Horlin, ``Mixed analog/digital beamforming for
  60 {GHz MIMO} frequency selective channels,'' in \emph{Proc.\ IEEE ICC},
  2010.

\bibitem{Rajagopal:mmWMobile}
S.~Rajagopal, S.~Abu-Surra, Z.~Pi, and F.~Khan, ``Antenna array design for
  multi-gbps mmwave mobile broadband communication,'' in \emph{Proc.\ IEEE
  Globecom}, 2011.

\bibitem{Huang:2008:MWA:1524107}
K.-C. Huang and D.~J. Edwards, \emph{Millimetre Wave Antennas for Gigabit
  Wireless Communications: A Practical Guide to Design and Analysis in a System
  Context}.\hskip 1em plus 0.5em minus 0.4em\relax Wiley Publishing, 2008.

\bibitem{Rusek:13}
F.~Rusek, D.~Persson, B.~K. Lau, E.~Larsson, T.~Marzetta, O.~Edfors, and
  F.~Tufvesson, ``Scaling up {MIMO}: Opportunities and challenges with very
  large arrays,'' \emph{IEEE Signal Process. Mag.}, vol.~30, no.~1, pp. 40--60,
  2013.

\bibitem{Rappaport:02}
T.~S. Rappaport, \emph{Wireless Communications: Principles and Practice},
  2nd~ed.\hskip 1em plus 0.5em minus 0.4em\relax Upper Saddle River, NJ:
  Prentice Hall, 2002.

\bibitem{Rappaport:12-28G}
Y.~Azar, G.~N. Wong, K.~Wang, R.~Mayzus, J.~K. Schulz, H.~Zhao, F.~Gutierrez,
  D.~Hwang, and T.~S. Rappaport, ``28 {GHz} propagation measurements for
  outdoor cellular communications using steerable beam antennas in {N}ew {Y}ork
  {C}ity,'' in \emph{Proc.\ IEEE ICC}, 2013.

\bibitem{Rappaport:28NYCPenetrationLoss}
H.~Zhao, R.~Mayzus, S.~Sun, M.~Samimi, J.~K. Schulz, Y.~Azar, K.~Wang, G.~N.
  Wong, F.~Gutierrez, and T.~S. Rappaport, ``28 {GHz} millimeter wave cellular
  communication measurements for reflection and penetration loss in and around
  buildings in {N}ew {Y}ork {C}ity,'' in \emph{Proc. IEEE ICC}, 2013.

\bibitem{Samimi:AoAD}
M.~K. Samimi, K.~Wang, Y.~Azar, G.~N. Wong, R.~Mayzus, H.~Zhao, J.~K. Schulz,
  S.~Sun, F.~Gutierrez, and T.~S. Rappaport, ``28 {GHz} angle of arrival and
  angle of departure analysis for outdoor cellular communications using
  steerable beam antennas in {N}ew {Y}ork {C}ity,'' in \emph{Proc. IEEE VTC},
  2013.

\bibitem{rappaportmillimeter}
T.~S. Rappaport, S.~Sun, R.~Mayzus, H.~Zhao, Y.~Azar, K.~Wang, G.~N. Wong,
  J.~K. Schulz, M.~Samimi, and F.~Gutierrez, ``Millimeter wave mobile
  communications for {5G} cellular: {I}t will work!'' \emph{IEEE Access},
  vol.~1, pp. 335--349, May 2013.

\bibitem{Rappaport:13-smallCell}
T.~S. Rappaport, G.~MacCartney, S.~Sun, and S.~Niu, ``Wideband millimeter-wave
  propagation measurements and models for small cell, peer-to-peer, and
  backhaul wireless communications,'' Submitted to \emph{IEEE JSAC Special
  Issue on 5G Cellular}, Dec. 2013.

\bibitem{Sun-Beam:13}
S.~Sun and T.~S. Rappaport, ``Multi-beam antenna combining for 28~{GHz}
  cellular link improvement in urban environments,'' in \emph{Proc. IEEE
  Globecom}, Dec. 2013.

\bibitem{3GPP36.814}
3GPP, ``{Further advancements for {E-UTRA} physical layer aspects},'' TR 36.814
  (release 9), 2010.

\bibitem{ITU-M.2134}
ITU, ``M.2134: {R}equirements related to technical performance for
  {IMT-A}dvanced radio interfaces,'' Technical Report, 2009.

\bibitem{AkLiuRanEr:13-GC}
M.~R. Akdeniz, Y.~Liu, S.~Rangan, and E.~Erkip, ``Millimeter wave picocellular
  system evaluation for urban deployments,'' in \emph{Globecom Workshops, 2013
  IEEE}, Dec. 2013.

\bibitem{Zwick05}
T.~Zwick, T.~Beukema, and H.~Nam, ``Wideband channel sounder with measurements
  and model for the 60 {GHz} indoor radio channel,'' \emph{IEEE Trans.\
  Vehicular Technology}, vol.~54, no.~4, pp. 1266 -- 1277, July 2005.

\bibitem{Giannetti:99}
F.~Giannetti, M.~Luise, and R.~Reggiannini, ``{Mobile and personal
  communications in 60 GHz band: A survey},'' \emph{Wirelesss Personal
  Comunications}, vol.~10, pp. 207 -- 243, 1999.

\bibitem{Anderson04}
C.~R. Anderson and T.~S. Rappaport, ``{In-building wideband partition loss
  measurements at 2.5 and 60 GHz},'' \emph{IEEE Trans. Wireless Comm.}, vol.~3,
  no.~3, pp. 922 -- 928, May 2004.

\bibitem{Smulders}
P.~Smulders and A.~Wagemans, ``{Wideband indoor radio propagation measurements
  at 58 GHz},'' \emph{Electronics Letters}, vol.~28, no.~13, pp. 1270 --1272,
  June 1992.

\bibitem{Manabe}
T.~Manabe, Y.~Miura, and T.~Ihara, ``{Effects of antenna directivity and
  polarization on indoor multipath propagation characteristics at 60 GHz},''
  \emph{IEEE J. Sel. Areas Comm.}, vol.~14, no.~3, pp. 441 --448, April 1996.

\bibitem{ben2011millimeter}
E.~Ben-Dor, T.~S. Rappaport, Y.~Qiao, and S.~J. Lauffenburger,
  ``Millimeter-wave {60 GH}z outdoor and vehicle {AOA} propagation measurements
  using a broadband channel sounder,'' in \emph{Proc.\ IEEE Globecom}, 2011,
  pp. 1--6.

\bibitem{ted2}
H.~Xu, V.~Kukshya, and T.~S. Rappaport, ``{Spatial and temporal characteristics
  of 60 GHz indoor channel},'' \emph{IEEE J. Sel. Areas Comm.}, vol.~20, no.~3,
  pp. 620 -- 630, April 2002.

\bibitem{ZhangMadhow1}
H.~Zhang, S.~Venkateswaran, and U.~Madhow, ``{Channel modeling and {MIMO}
  capacity for outdoor millimeter wave links},'' in \emph{Proc. IEEE WCNC},
  April 2010.

\bibitem{AkoumAyaHeath:12}
S.~Akoum, O.~E. Ayach, and R.~W. {Heath, Jr.}, ``Coverage and capacity in
  {mmWave} cellular systems,'' in \emph{Proc. of Asilomar Conf. on Signals,
  Syst. \& Computers}, Pacific Grove, CA, Nov. 2012.

\bibitem{ElrefShak:97}
A.~Elrefaie and M.~Shakouri, ``Propagation measurements at 28 {GH}z for
  coverage evaluation of local multipoint distribution service,'' \emph{Proc.\
  Wireless Communications Conference}, pp. 12--17, Aug. 1997.

\bibitem{SeiArn:95}
S.~Seidel and H.~Arnold, ``Propagation measurements at 28 {GH}z to investigate
  the performance of {L}ocal {M}ultipoint {D}istribution {S}ervice ({LMDS}),''
  \emph{Proc.\ Wireless Communications Conference}, pp. 754--757, Nov. 1995.

\bibitem{Rappaport:13-BBmmW}
T.~S. Rappaport, F.~Gutierrez, E.~Ben-Dor, J.~N. Murdock, Y.~Qiao, and J.~I.
  Tamir, ``Broadband millimeter-wave propagation measurements and models using
  adaptive-beam antennas for outdoor urban cellular communications,''
  \emph{IEEE Trans.\ Antennas and Propagation}, vol.~61, no.~4, pp. 1850--1859,
  2013.

\bibitem{Rappaport38:12}
T.~S. Rappaport, E.~Ben-Dor, J.~Murdock, and Y.~Qiao, ``38~{GH}z and 60~{GH}z
  angle-dependent propagation for cellular and peer-to-peer wireless
  communications,'' \emph{Proc.\ IEEE ICC}, pp. 4568--4573, Jun. 2012.

\bibitem{ted:rww12}
T.~S. Rappaport, E.~Ben-Dor, J.~Murdock, Y.~Qiao, and J.~Tamir, ``Cellular
  broadband millimeter wave propagation and angle of arrival for adaptive beam
  steering systems,'' in \emph{Proc. IEEE RWS (Invited)}, Jan. 2012.

\bibitem{ted:wcnc12}
J.~Murdock, E.~Ben-Dor, Y.~Qiao, J.~Tamir, and T.~S. Rappaport, ``A 38~{GH}z
  cellular outage study for an urban outdoor campus environment,'' in
  \emph{Proc. IEEE WCNC}, April 2012.

\bibitem{Rappaport-72GHz:13}
G.~R. {MacCartney, Jr.}, J.~Zhang, S.~Nie, and T.~S. Rappaport, ``Path loss
  models for {5G} millimeter wave propagation channels in urban microcells,''
  in \emph{Proc. IEEE Globecom}, Dec. 2013.

\bibitem{wells:09}
J.~A. Wells, ``Faster than fiber: The future of multi-{G}b/s wireless,''
  \emph{IEEE Microwave Magazine}, vol.~10, no.~3, pp. 104--112, May 2009.

\bibitem{Bishop:06}
C.~M. Bishop, \emph{Pattern Recognition and Machine Learning}, ser. Information
  Science and Statistics.\hskip 1em plus 0.5em minus 0.4em\relax New York, NY:
  Springer, 2006.

\bibitem{winner2:07}
P.~Ky{\"o}sti, J.~Meinil{\"a}, L.~Hentil{\"a}, X.~Zhao, T.~J{\"a}ms{\"a},
  M.~Narandzic, M.~Milojevic, C.~Schneider, A.~Hong, J.~Ylitalo, V.-M. Holappa,
  M.~Alatossava, R.~Bultitude, Y.~de~Jong, and T.~Rautiainen, ``{WINNER II}
  channel models part {II} radio channel measurement and analysis results,''
  IST-4-027756 {WINNER II} D1.1.2 V1.0, 2007.

\bibitem{Heath:14arXiv}
T.~Bai and R.~W. {Heath, Jr.}, ``Coverage and rate analysis for millimeter wave
  cellular networks,'' arXiv:1402.6430v2 [cs.IT]., Mar. 2014.

\bibitem{bai2013analysis}
T.~Bai, R.~Vaze, and R.~W. {Heath Jr.}, ``Analysis of blockage effects on urban
  cellular networks,'' \emph{arXiv preprint arXiv:1309.4141}, Sep. 2013.

\bibitem{Nie28-73SOS:14}
S.~Nie, G.~R. {MacCartney, Jr}., S.~Sun, and T.~S. Rappaport, ``28 {GH}z and 73
  {GH}z signal outage study for millimeter wave cellular and backhaul
  communications,'' in \emph{Proc.\ IEEE ICC}, Jun. 2014, to appear.

\bibitem{TseV:07}
D.~Tse and P.~Viswanath, \emph{{Fundamentals of Wireless Communication}}.\hskip
  1em plus 0.5em minus 0.4em\relax Cambridge University Press, 2007.

\bibitem{Heath:partialBF}
A.~Alkhateeb, O.~E. Ayach, G.~Leus, and R.~W. {Heath, Jr.}, ``Hybrid precoding
  for millimeter wave cellular systems with partial channel knowledge,'' in
  \emph{Proc.\ Information Theory and Applications Workshop (ITA)}, Feb. 2013.

\bibitem{ParkZim:02}
D.~Parker and D.~Z. Zimmermann, ``{Phased arrays-part I: Theory and
  architecture},'' \emph{IEEE Trans.\ Microw.\ Theory Tech.}, vol.~50, no.~3,
  pp. 678--687, Mar. 2002.

\bibitem{KohReb:07}
K.-J. Koh and G.~M. Rebeiz, ``{0.13-$\mu$m CMOS phase shifters for X-, Ku- and
  K-band phased arrays},'' \emph{IEEE J. Solid-State Circuts}, vol.~42, no.~11,
  pp. 2535--2546, Nov. 2007.

\bibitem{KohReb:09}
------, ``A millimeter-wave (40$-$45 {GH}z) 16-element phased-array transmitter
  in 0.18-$\mu$m {SiGe} {BiCMOS} technology,'' \emph{IEEE J. Solid-State
  Circuts}, vol.~44, no.~5, pp. 1498--1509, May 2009.

\bibitem{Crane-Patent:88}
P.~E. Crane, ``{Phased array scanning system},'' United States Patent
  4,731,614, filed Aug 11, 1986, issued Mar.\ 15, 1988.

\bibitem{RamBaRe:98}
S.~Raman, N.~S. Barker, and G.~M. Rebeiz, ``{A W-band dielectriclens-based
  integrated monopulse radar receive},'' \emph{IEEE Trans.\ Microw.\ Theory
  Tech.}, vol.~46, no.~12, pp. 2308--2316, Dec. 1998.

\bibitem{GuanHaHa:04}
X.~Guan, H.~Hashemi, and A.~Hajimiri, ``{A fully integrated 24-GHz
  eight-element phased-array receiver in silicon},'' \emph{IEEE J. Solid-State
  Circuts}, vol.~39, no.~12, pp. 2311--2320, Dec. 2004.

\bibitem{Lozano:07}
A.~Lozano, ``Long-term transmit beamforming for wireless multicasting,'' in
  \emph{Proc.\ ICASSP}, vol.~3, 2007, pp. III--417--III--420.

\bibitem{kermoal2002stochastic}
J.~P. Kermoal, L.~Schumacher, K.~I. Pedersen, P.~E. Mogensen, and
  F.~Frederiksen, ``A stochastic {MIMO} radio channel model with experimental
  validation,'' \emph{IEEE J. Sel. Areas Comm.}, vol.~20, no.~6, pp.
  1211--1226, 2002.

\bibitem{mcnamara2002spatial}
D.~McNamara, M.~Beach, and P.~Fletcher, ``Spatial correlation in indoor {MIMO}
  channels,'' in \emph{Proc.\ IEEE PIMRC}, vol.~1, 2002, pp. 290--294.

\bibitem{Z3}
S.~Pinel, S.~Sarkar, P.~Sen, B.~Perumana, D.~Yeh, D.~Dawn, and J.~Laskar, ``{A
  90nm {CMOS} 60 {GH}z Radio},'' in \emph{Proc.\ IEEE International Solid-State
  Circuits Conference}, 2008.

\bibitem{Z4}
C.~Marcu, D.~Chowdhury, C.~Thakkar, J.-D. Park, L.-K. Kong, M.~Tabesh, Y.~Wang,
  B.~Afshar, A.~Gupta, A.~Arbabian, , S.~Gambini, R.~Zamani, E.~Alon, and
  A.~Niknejad, ``A 90 nm {CMOS} low-power 60 {GH}z transceiver with integrated
  baseband circuitry,'' \emph{IEEE J. Solid-State Circuts}, vol.~44, pp.
  3434--3447, 2009.

\bibitem{MogEtAl:07}
P.~Mogensen, W.~Na, I.~Z. Kov{\'a}cs, F.~Frederiksen, A.~Pokhariyal, K.~I.
  Pedersen, T.~Kolding, K.~Hugl, and M.~Kuusela, ``{LTE} capacity compared to
  the {S}hannon bound,'' in \emph{Proc. IEEE VTC}, 2007.

\bibitem{ikuno2011novel}
J.~C. Ikuno, C.~Mehlfuhrer, and M.~Rupp, ``A novel link error prediction model
  for {OFDM} systems with {HARQ},'' in \emph{Proc. IEEE ICC}, 2011.

\end{thebibliography}

\end{document}